\RequirePackage{rotating}
\documentclass[fleqn,usenatbib]{mnras}

\usepackage[T1]{fontenc}

\DeclareRobustCommand{\VAN}[3]{#2}
\let\VANthebibliography\thebibliography
\def\thebibliography{\DeclareRobustCommand{\VAN}[3]{##3}\VANthebibliography}

\usepackage{graphicx}	
\usepackage{amsmath}	
\usepackage{amssymb}	
\usepackage{newtxtext,newtxmath}
\usepackage{multirow}
\usepackage{lscape}
\usepackage{makecell}
\usepackage{subfigure}
\usepackage{float}
\usepackage{newtxtext,newtxmath}
\usepackage{lineno}
\setcitestyle{notesep={ }}
\makeatletter
\newcommand{\labtext}[2]{%
  \@bsphack
  \csname phantomsection\endcsname 
  \def\@currentlabel{#1}{\label{#2}}%
  \@esphack
}
\newcommand\T{\rule{0pt}{2.7ex}}
\newcommand\Tshort{\rule{0pt}{1.65ex}}

\setcitestyle{notesep={}}

\makeatother

\title[MWL Variability Analysis of Blazars]{Multiwavelength Variability Analysis of \textit{Fermi}-LAT Blazars}

\author[P. Pe\~nil et al.]{
P. Pe\~nil,$^{1}$\thanks{E-mail: ppenil@clemson.edu}
J. Otero-Santos,$^{2,3,4}$\thanks{E-mail: joteros@iaa.es}
M. Ajello, $^{1}$\thanks{E-mail: majello@clemson.edu}
S. Buson,$^{5}$\thanks{E-mail: sara.buson@uni-wuerzburg.de}
A. Dom\'inguez,$^{6}$
L. Marcotulli,$^{7}$
N. Torres$-$Alb\`a,$^{1}$
\newauthor
J. Becerra Gonz\'alez,$^{2,3}$
J.A. Acosta-Pulido,$^{2,3}$
\\
$^{1}$Department of Physics and Astronomy, Clemson University, Kinard Lab of Physics, Clemson, SC 29634-0978, USA\\
$^{2}$Instituto de Astrof\'isica de Canarias (IAC), E-38200 La Laguna, Tenerife, Spain\\
$^{3}$Universidad de La Laguna (ULL), Departamento de Astrof\'isica, E-38206 La Laguna, Tenerife, Spain\\
$^{4}$Instituto de Astrof\'isica de Andalucía (CSIC), Glorieta de la Astronomía s/n, 18008 Granada, Spain\\
$^{5}$Julius-Maximilians-Universit\"at, 97070, W\"urzburg, Germany\\
$^{6}$IPARCOS and Department of EMFTEL, Universidad Complutense de Madrid, E-28040 Madrid, Spain\\
$^{7}$ Department of Physics, Yale University, 52 Hillhouse Avenue, New Haven, CT 06511, USA\\
}

\date{Accepted  2024 February 14. Received 2024 February 9; in original form 2023 October 13}

\pubyear{2023}

\begin{document}
\label{firstpage}
\pagerange{\pageref{firstpage}--\pageref{lastpage}}
\maketitle
\begin{abstract}
Blazars present highly variable $\gamma$-ray emission. This variability, which can range from a few minutes to several years, is also observed at other wavelengths across the entire electromagnetic spectrum. We make use of the first 12 years of data from the \textit{Fermi} Large Area Telescope (LAT), complemented with multiwavelength (MWL) archival data from different observatories and facilities in radio, infrared and optical bands, to study the possible periodic emission from 19 blazars previously claimed as periodic candidates. A periodicity analysis is performed with a pipeline for periodicity searches. Moreover, we study the cross-correlations between the $\gamma$-ray and MWL light curves. Additionally, we use the fractional variability and the structure function to evaluate the variability timescales. We find five blazars showing hints of periodic modulation with $\geq$3.0$\sigma$ ($\approx$0$\sigma$ post-trials), with periods ranging from 1.2 to 4 years, both in their $\gamma$-ray and MWL emission. The results provide clues for understanding the physical mechanisms generating the observed periodicity.
\end{abstract}


\section{Introduction} \label{sec:intro}
Supermassive black holes (SMBHs) are found in the centers of almost all galaxies. Accretion of matter onto SMBHs powers some of the most luminous sources in the Universe known as active galactic nuclei \citep[AGNs, ][]{urry1995}. Some of these AGNs produce powerful, highly collimated relativistic jets. When these jets are aligned with the line of view of the observer, the jets' emission is relativistically boosted from radio to $\gamma$ rays \citep{ghisellini1998}, and these AGNs are classified as blazars. One of the main features of blazars is their strong variability at all wavelengths and at different timescales \citep{fan2005}. 

This variability is generally interpreted as arising due to stochastic and unpredictable processes \citep[e.g. ][]{ruan2012}. However, several studies have claimed the detection of quasi-periodic signals coming from blazars \citep[see, e.g. ][]{ackermann_pg1553, sandrinelli2016, penil_2020}. This latter scenario has grown in interest over the last decades due to the substantial implications derived from periodic emissions in these objects. It can provide information about the nature of the source and the physical processes involved in the most violent environments. Quasi-periodic oscillations in blazars have been interpreted, for instance, in the framework of a binary SMBH system \citep[e.g. ][]{sobacchi2017}. This scenario has been proposed especially for the two best candidates: PG~1553+113 \citep{tavani2018} and OJ~287 \citep{sillanpaa1988,valtonen2011}. Nevertheless, periodicity studies and detections of periodic oscillations in blazars are still a controversial topic \citep[see, e.g. ][]{covino2019, yang_carma}. The impact of the noise in data can generate stochastic uncertainty in the periodicity search, which can provoke the detection of fake periodicity \citep[e.g., ][]{vaughan_criticism, liu_pg1302}.

Alternative explanations for periodic emission in blazars include geometrical effects like, e.g., the lighthouse effects \citep[e.g., ][]{camenzind1992}, periodic precession of the jet, helical jets \citep[e.g., ][]{rieger2004}, helical structures in the magnetic field \citep{raiteri2013}, periodic shocks, instabilities propagating along the jet \citep[e.g., ][]{mohan2015}, or periodic modulations of the accretion flow \citep[e.g., ][]{dong2020}.

The recent study from \citet{penil_2020} (P20 hereafter) presents a systematic search for periodicity in blazars observed by the \textit{Fermi} Large Area Telescope \citep[LAT, ][]{fermi_lat}. As a result, a sample of 24 (out of 351 analyzed, more details in P20) blazars present hints of periodic emission with confidence between 2$\sigma$ and $\sim$4$\sigma$ (pre-trials). In \citet[][, hereafter P22]{penil_2022}, these 24 blazars were reanalyzed with three more years of \textit{Fermi}-LAT observations, for a total of 12 years.  
Here, we analyze the low-significance 19 candidates with a period in $\gamma$-ray of $<$4$\sigma$ (pre-trials) reported in the latter study as the first publication of two papers, extending the analysis to the  MWL data collected for the 24 periodic candidates. 
We report the results for PG 1553+113, PKS 2155-304,  S5 0716+714, OJ 014, and PKS 0454-304 in a separate study \citep[][]{penil_mwl_pg1553} since these sources stand out from the sample in showing higher significance in comparison with the rest of the sample for the quasi-periodicity (with pre-trials significance of $\approx$4$\sigma$ in the period observed in $\gamma$ rays).

The paper is structured as follows. In Section \ref{sec:fermidata}, we introduce the sample of sources analyzed in this work and the \textit{Fermi}-LAT and MWL data collected. In Section \ref{sec:methodology}, we describe the methodology and analysis tools. In Section \ref{sec:results}, the results of the analysis are presented. Section \ref{sec:discussion} provides a possible interpretation of the results. Section \ref{sec:conclusions} summarizes the main results and conclusions. Finally, we also include an appendix in Section \ref{sec:appendix} where all the MWL LCs of the blazars are presented. 

\section{Blazar Sample} \label{sec:fermidata}
\subsection{$\gamma$-ray Sample} \label{sec:sample}
In P20, the analysis of 351 blazars detected by \textit{Fermi}-LAT in 9 years of data (0.1~GeV$<$E$<$800~GeV) was carried out. This analysis was performed with a pipeline consisting of several methods (see $\S$\ref{sec:methodology}), leading to the discovery of 24 objects with evidence of periodic emission with confidence between 2$\sigma$ and $\sim$4$\sigma$ (computing the median significance across the analysis methods to sort the blazars). These values refer to the ``local significance'' since no trial-factor corrections were included in P20. Therefore, no ``global significance'' was reported. These 24 blazars are reanalyzed in P22, updating the sample with 12 years of \textit{Fermi}-LAT data at energies $>$0.1 GeV, where the evidence of periodicity is confirmed for 5 of them with a significance (pre-trials) of $>$3$\sigma$ (again, using the same pipeline and the median significance across the analysis methods). The ``global significance'' of these blazars is $\gtrsim$2.0$\sigma$. 

The remaining 19 objects studied here reported a local significance (pre-trials) between 1.0$\sigma$ and $\sim$3.0$\sigma$. Their global significance is $\approx$ 0$\sigma$ (see details in $\S$ \ref{sec:periodicty_results}). This global significance does not allow us to claim any periodicity. However, we will search for the same periods in the MWL bands observed in $\gamma$ rays. These results allow us to select the most promising in order to follow them up to confirm or rule out the periods inferred.

\begin{figure}
	\centering
	\includegraphics[width=\columnwidth]{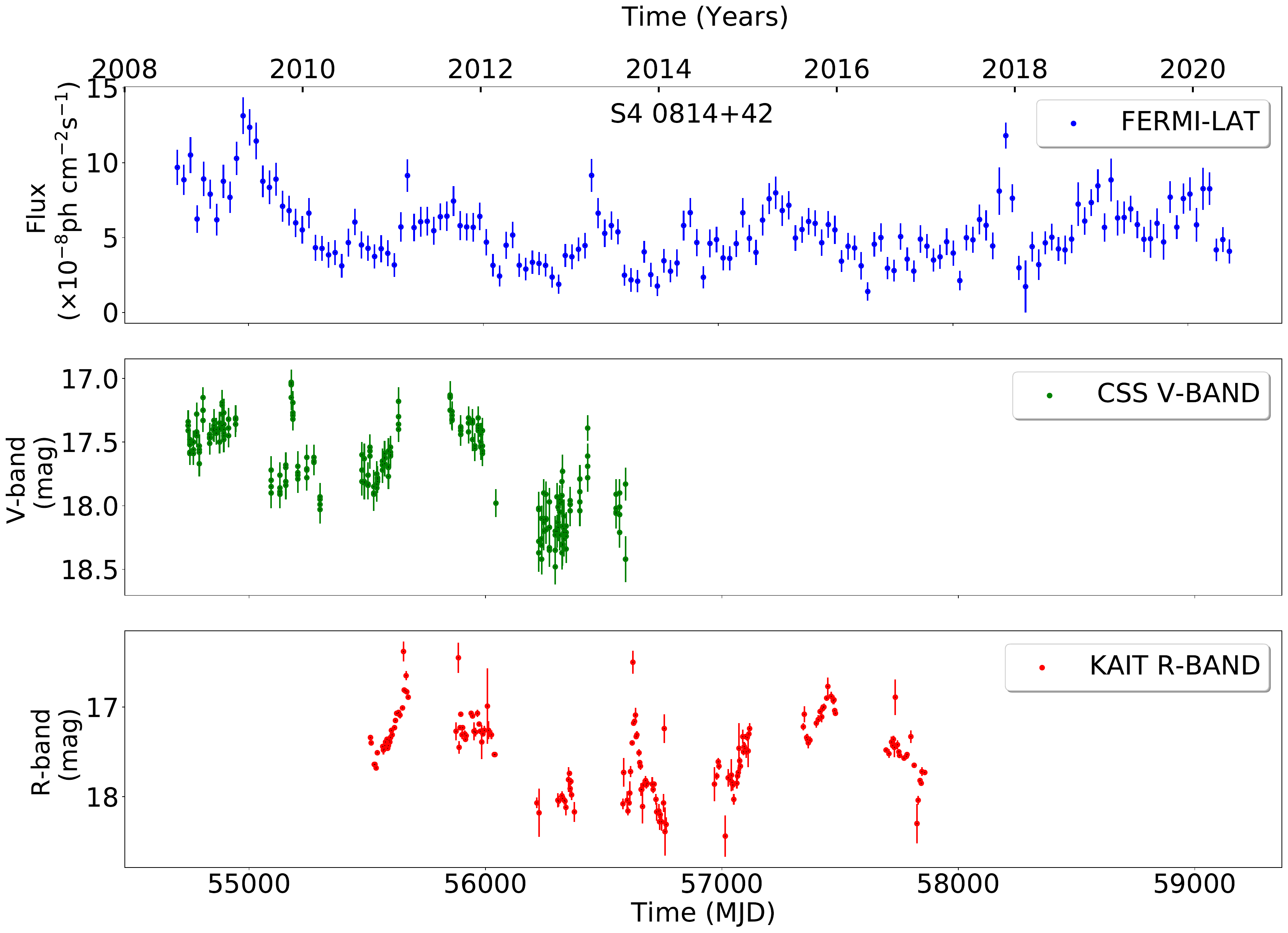}
	\caption{MWL light curves of S4 0814+42. From top to bottom: \textit{Fermi}-LAT (E $>$ 0.1 GeV), V-band (CCS), and R-band (KAIT) light curves.}
	\label{fig:mwl_lc}
\end{figure}

The list of the selected blazars can be found in Table \ref{tab:candidates_list}. We use the \textit{Fermi}-LAT light curves (LCs) from P22, complemented with MWL data from several observatories listed in Section \ref{sec:wave_data}. An example of the LCs used in this work is shown in Figure \ref{fig:mwl_lc} for S4~0814+42. 

\subsection{Multiwavelength Data} \label{sec:wave_data}
For the MWL analysis, we employ archival data from different databases and observatories, covering most of the electromagnetic spectrum. In the X-ray bands, we use data from the Neil Gehrels Swift Observatory, specifically {\it Swift}-XRT (X-Ray Telescope)\footnote{\url{https://www.swift.ac.uk/analysis/xrt/}}. The raw count-rate data were taken from the automatic processing of \cite{stroh_monitoring}\footnote{\url{http://www.swift.psu.edu/monitoring/}}. Hardness ratio estimations (HR, ratio between soft (0.3$-$2.0 keV) and hard (2.0$-$10.0 keV) X-rays), also by \cite{stroh_monitoring}, were used in conjunction with the {\it Swift}-XRT detector response to estimate a photon index, $\alpha$, under the assumption that blazar emission in X-rays can be represented by a simple power-law (PL) A(E)=KE$^{-\alpha}$ \citep[][]{ghisellini_canonical_blazars, middei_x_ray_cat}. The count rate and HR were then used to estimate a flux at each given epoch. We use data from \textit{Swift}-UVOT (Ultraviolet and Optical Telescope)\footnote{\url{https://www.swift.ac.uk/about/instruments.php}} for the filters `uvw2' (1928~\AA), `uvm2' (2246~\AA) and `uvw1' (2600~\AA)\footnote{\url{https://www.swift.ac.uk/analysis/uvot/filters.php}}. We perform a data reduction of all available {\it Swift}-UVOT archival observations to provide the optical-to-UV LCs via the standard pipeline, detailed in \citet{poole_swift_analysis}. For all UVOT filters, the source regions are selected as circles of 5$\arcsec$, centered on the source. The background is defined as a circle of 30$\arcsec$ away from the source to avoid contamination. The task \texttt{uvotsource} is employed to extract the magnitudes, which are then corrected for Galactic extinction according to the recommendations in \citet{roming_uvot_analysis}. Finally, the fluxes are derived using the standard zero points listed in \citet{breeveld_calibration_swift}. 

The optical data are obtained from KAIT \citep[Katzman Automatic Imaging Telescope, R band, ][]{li2003}\footnote{\url{http://herculesii.astro.berkeley.edu/kait/agn/}}, CSS \citep[Catalina Sky Survey, V band, ][]{drake2009}\footnote{\url{http://nesssi.cacr.caltech.edu/DataRelease/}}, and ASAS-SN \citep[All-Sky Automated Survey for Supernovae, V band, ][]{shappee2014,kochanek2017}\footnote{\url{http://www.astronomy.ohio-state.edu/asassn/index.shtml}}. Data from the Tuorla blazar monitoring program\footnote{\url{http://users.utu.fi/kani/1m/}} \citep[R band, ][]{takalo_tuorla} are also included, using the data from \cite{tuorla_data}. We also employ the optical V- and R-band data from the American Association of Variable Star Observers (AAVSO)\footnote{\url{http://https://www.aavso.org/data-download/}} International Database. We use data from SMARTS \citep[Small and Moderate Aperture Research Telescope System, ][]{bonning2012}\footnote{\url{http://www.astro.yale.edu/smarts/glast/home.php}} in the optical B, V, and R bands, and near-infrared (IR) J and K bands. Finally, we also retrieve data from the Astronomy \& Steward Observatory\footnote{\url{http://james.as.arizona.edu/~psmith/Fermi/}} \citep{smith2009}, with optical V- and R-band observations. For the analysis, we combine all the data from different observatories in the V and R bands, denoted in the tables and figures as "V-band" and "R-band". Different calibrations from different observatories can introduce offsets that affect the data. To evaluate and correct for the presence of such an effect, we compare all the simultaneous data from the different databases. If the simultaneous data show a systematic offset, the mean difference between each simultaneous pair of measurements is calculated. This estimation is used as the mean offset to correct the data from different observatories and ensure their compatibility. 

Since we are interested in studying the long-term evolution of these data series, this scaling is irrelevant to the analysis performed in this work. Some blazars use non-calibrated V- and R-band data from the Steward Observatory. These data can be used for analyzing long-term time series. However, they are not combined with the rest of the LCs of the corresponding band. These data are denoted in the tables as ``Steward-V'' and ``Steward-R'', respectively. 

Moreover, we also include radio data in the 15 GHz band for those sources observed by the 40m radio telescope from Owens Valley Radio Observatory (OVRO) as part of their blazar monitoring program \citep{richards2011}\footnote{\url{https://www.astro.caltech.edu/ovroblazars/}}. These data extend for longer than 12 years, to June 2020, when monitoring ceased, and the data were made publicly available. Radio data are available for 5 of the 19 sources studied in this work. Data with a signal-to-noise ratio lower than 3 are considered non-detection and excluded from the analysis to avoid such data. To perform reliable periodicity and correlation analyses, we need to balance reducing the uncertainties introduced by the noisiest observations and maintaining sufficient sampling and time coverage in the LCs. Otherwise, the analysis could become unreliable or result in a poor estimation of the period/correlation \citep[][]{otero2023}.

As we are interested in yearly-scale variations, we use 28-day bins for the MWL LC periodicity analysis, matching the bin size from the \textit{Fermi}-LAT $\gamma$-ray LCs. The data is organized into 28-day bins, a crucial step that allows us to explore variations beyond the short-term fluctuations, including those occurring within a single day or week. Our binning method relies on the median value within each bin, which has proven to be effective in managing data with erratic noise, as demonstrated in previous studies \citep[e.g., ][]{bindu_binning, negi_binning}. This approach strikes a balance between making the analysis computationally feasible while still maintaining sensitivity to longer-term variations, which may extend over the course of a year. On the other hand, the correlation with the $\gamma$-ray LCs (to have an accurate estimation of the time lags) and the variability analyses (the exception is the fractional variability, see $\S$\ref{sec:variability} for details) are performed using the original sampling of the LCs. 

\begin{table*}
\begin{center}
\caption{List of the blazars analyzed here in descending order of the median significance (pre-trials) of the quasi-periodic hypothesis, estimated from the results presented in Table \ref{tab:periodicity_results}. Note that this median significance does not have an actual statistical meaning; it is used as an arbitrary way of combining all of the significance for a given source, sorting the candidates, and comparing with the results obtained in P22 for the $\gamma$-rays. The blazars are characterized by their \textit{Fermi}-LAT name, coordinates, blazar class, the blazar type according to the frequency of the synchrotron peak (LSP: low synchrotron peaked, ISP: intermediate synchrotron peaked, HSP: high synchrotron peaked), redshift, association name. Additionally, we include the average period (in years) with the uncertainty and the local significance obtained in P22 (as the average of the periodicity) and the corresponding global significance estimated in P22. Note that some sources have two significant periods (organized by the amplitude of the peak), denoted by $\star$.}
 \label{tab:candidates_list}
\begin{tabular}{cccccccccc} \hline
(1) & (2) & (3) & (4) & (5) & (6) & (7) & (8) & (9) & (10) \\ 
\multirow{2}{*}{4FGL Source Name} & \multirow{2}{*}{RA(J2000)}  & \multirow{2}{*}{DEC(J2000)} & \multirow{2}{*}{Type}  & \multirow{2}{*}{Subtype} & \multirow{2}{*}{Redshift} & \multirow{2}{*}{Association Name} & Period in $\gamma$ rays & Local & Global  \\ 
 &   &  &  &  &  &  & [yrs] & (S/N) &  (S/N)  \\ \hline 
        J0043.8+3425 & 10.96782 & 34.42687 & fsrq & LSP & 0.966 & GB6 J0043+3426 & 1.9$\pm$0.2 & 2.8$\sigma$ & $\sim$0$\sigma$ \\
        J0521.7+2113 & 80.44379 & 21.21369 & bll & HSP & 0.108 & TXS 0518+211 & 3.1$\pm$0.4 & 2.8$\sigma$ & $\sim$0$\sigma$ \\
        J0449.4$-$4350 & 72.36042 & $-$43.83719 & bll & HSP & 0.205 & PKS 0447$-$439 & 1.9$\pm$0.2  & 2.7$\sigma$ & $\sim$0$\sigma$ \\ 
	J0252.8$-$2218 & 43.20377 & $-$22.32386 & fsrq & LSP & 1.419 & PKS 0250$-$225 & 1.2$\pm$0.1  & 2.7$\sigma$ & $\sim$0$\sigma$ \\
        J1146.8+3958 & 176.73987 & 39.96861 & fsrq & LSP & 1.089 & S4 1144+40 & 3.3$\pm$0.5 & 2.3$\sigma$ & $\sim$0$\sigma$ \\
        J0303.4$-$2407 & 45.86259 & $-$24.12074 & bll & HSP & 0.266 & PKS 0301$-$243 & 2.1$\pm$0.2  & 2.2$\sigma$ & $\sim$0$\sigma$ \\
        J0428.6$-$3756 & 67.17261 & $-$37.94081 & bll & LSP & 1.11 & PKS 0426$-$380 & 3.6$\pm$0.5 & 2.1$\sigma$ & $\sim$0$\sigma$\\ 
		J1248.2+5820$\star$ & 192.07728 & 58.34622 & bll & ISP & -- & PG 1246+586 & \makecell{\\[0.1pt]2.1$\pm$0.2 \\ 1.4$\pm$0.1} & \makecell{\\[0.1pt]1.9$\sigma$  \\ 1.7$\sigma$} & \makecell{\\[0.1pt]$\sim$0$\sigma$ \\ $\sim$0$\sigma$} \\	
  	J2258.0$-$2759$\star$ & 344.50485 & $-$27.97588 & fsrq & LSP & 0.926 & PKS 2255$-$282 & \makecell{\\[0.1pt]2.8$\pm$0.3 \\ 1.4$\pm$0.1} & \makecell{\\[0.1pt]1.9$\sigma$ \\ 1.8$\sigma$} & \makecell{\\[0.1pt]$\sim$0$\sigma$ \\ $\sim$0$\sigma$} \\
	J1903.2+5541 & 285.80851 & 55.67557 & bll & LSP & -- & TXS 1902+556 & 3.3$\pm$0.3 & 1.8$\sigma$ & $\sim$0$\sigma$ \\
	J0818.2+4223 & 124.56174 & 42.38367 & bll & LSP &  0.530 & S4 0814+42 & 2.2$\pm$0.2 & 1.8$\sigma$ & $\sim$0$\sigma$ \\
 	J0211.2+1051 & 32.81532 & 10.85811 & bll & ISP & 0.2 & MG1 J021114+1051 & 2.9$\pm$0.3 & 1.8$\sigma$ & $\sim$0$\sigma$ \\	
	J0501.2$-$0157 & 75.30886 & $-$1.98359 & fsrq & LSP & 2.291 & S3 0458$-$02 & 3.8$\pm$0.6 & 1.8$\sigma$ & $\sim$0$\sigma$ \\
        J2056.2$-$4714$\star$ & 314.06768 & $-$47.23386 & fsrq & LSP & 1.489 & PKS 2052$-$47 & \makecell{\\[0.1pt]3.1$\pm$0.4 \\ 1.7$\pm$0.2} & \makecell{\\[0.1pt]1.7$\sigma$ \\ 2.1$\sigma$} & \makecell{\\[0.1pt]$\sim$0$\sigma$ \\ $\sim$0$\sigma$} \\
 	J1303.0+2435 & 195.75454 & 24.56873 & bll & LSP & 0.993 & MG2 J130304+2434 & 2.1$\pm$0.2 & 1.6$\sigma$ & $\sim$0$\sigma$ \\
 	J0102.8+5825 & 15.71134 & 58.41576 & fsrq & LSP & 0.644 & TXS 0059+581 & 4.0$\pm$0.6 & 1.4$\sigma$ & $\sim$0$\sigma$ \\
  	J1454.5+5124 & 223.63225 & 51.413868 & bll & ISP & -- & TXS 1452+516 & 2.1$\pm$0.3 & 1.2$\sigma$ & $\sim$0$\sigma$ \\
	J0210.7$-$5101 & 32.68952 & $-$51.01695 & fsrq & LSP & 1.003 & PKS 0208$-$512 & 3.8$\pm$0.5 & 1.1$\sigma$ & $\sim$0$\sigma$ \\ 
        J1649.4+5238 & 252.35208 & 52.58336 & bll & ISP & -- & 87GB 164812.2+524023 & 2.8$\pm$0.3 & 1.1$\sigma$ & $\sim$0$\sigma$ \\\hline 

\end{tabular}
\end{center}
\end{table*}

\section{Methodology}\label{sec:methodology}
\subsection{Periodicity}
The MWL study presented here is focused on two aspects: (1) periodicity and correlation search and (2) variability. The periodicity search is performed using the pipeline described in P20 and P22, to which we refer the reader for further details. This pipeline searches for periodic variability using multiple methods, which are listed below:

\begin{enumerate}
    \item Lomb-Scargle periodogram \citep[LSP, ][]{lomb_1976, scargle_1982}
    \item Generalized Lomb-Scargle periodogram \citep[GLSP, ][]{lomb_gen}
    \item Phase Dispersion Minimization \citep[PDM, ][]{pdm}
    \item Enhanced Discrete Fourier Transform with Welch's method \citep[DFT-Welch, ][]{welch}
    \item Weighted Wavelet Z-transform \citep[WWZ, ][]{wwz}
    \item Markov Chain Monte Carlo Sinusoidal Fitting \citep[MCMC Sine, ][]{emcee}.
\end{enumerate}

The selection of these methods for our analysis is based on their strong performance in handling unevenly sampled data, particularly LSP, GLSP, and WWZ, which were specifically developed to address this challenge. The LSP, GLSP, and DFT methods are all Fourier-transform-based, making them well-suited for detecting sinusoidal-like signals even in data with irregular sampling. WWZ, while also rooted in the Fourier transform, offers the additional advantage of decomposing the signal into both time and frequency domains, enabling an evaluation of the persistence of potential periods. PDM, in contrast, relies on the dispersion of LC data for various periods, making it a reliable choice for detecting non-sinusoidal periodic patterns, including repeating flares. The MCMC Sine fitting method stands apart from the others as it solely relies on the LC series and searches for sinusoidal signals without encountering the typical limitations associated with Fourier-transform methods, such as restricted frequency resolution, aliasing, and the presence of spurious peaks.

As demonstrated in prior studies P20, P22, and \cite{otero2023}, these methods can also exhibit different sensitivities to data gaps or red noise. By employing a combination of these methods, we aim to complement each other's strengths and weaknesses, ultimately providing more robust results when there is a consensus among multiple methods.

Another challenge in periodicity analysis, as highlighted in \cite{feigelson_2022}, pertains to non-stationary LCs. This means that estimating the power spectral density (PSD, a measurement of the power in a signal as a function of the frequency) cannot be uniformly applied to the entire time series because the source of the variability changes over time \citep{vaughan_2013}. To address this issue, we employ a detrending step. Detrending is a recommended preprocessing procedure \cite[e.g.,][]{detrend_welsh} to mitigate contamination that could otherwise lead to the incorrect inference of false periodicities \citep{mcquillan2013}. In our study, we opt for linear detrending, which can increase noise correlation, especially if the linear component is not effectively removed or if the original data contains non-random systematic patterns or fluctuations. However, we can rule out the latter case, as our blazar LCs are characterized by exhibiting red noise characteristics \cite[e.g.,][]{vaughan2003}.

The methods employed in our methodology are indeed effective in handling non-stationary LCs. Additionally, we take measures to ensure that the data becomes stationary after the detrending process. To achieve this, we utilize the augmented Dickey-Fuller test \citep[][]{dickey_fuller}. This test serves as a validation step to confirm the stationarity of the data \citep[][]{feigelson_arima}.

To perform the periodicity analysis, we perform a binning of 28 days in the MWL LC to be consistent with the $\gamma$-rays LCs (see P20 and P22). We search for long-term periodicity ($\sim$years), periods in the range of [1-6] years. 

\begin{figure}
\includegraphics[width=\columnwidth]{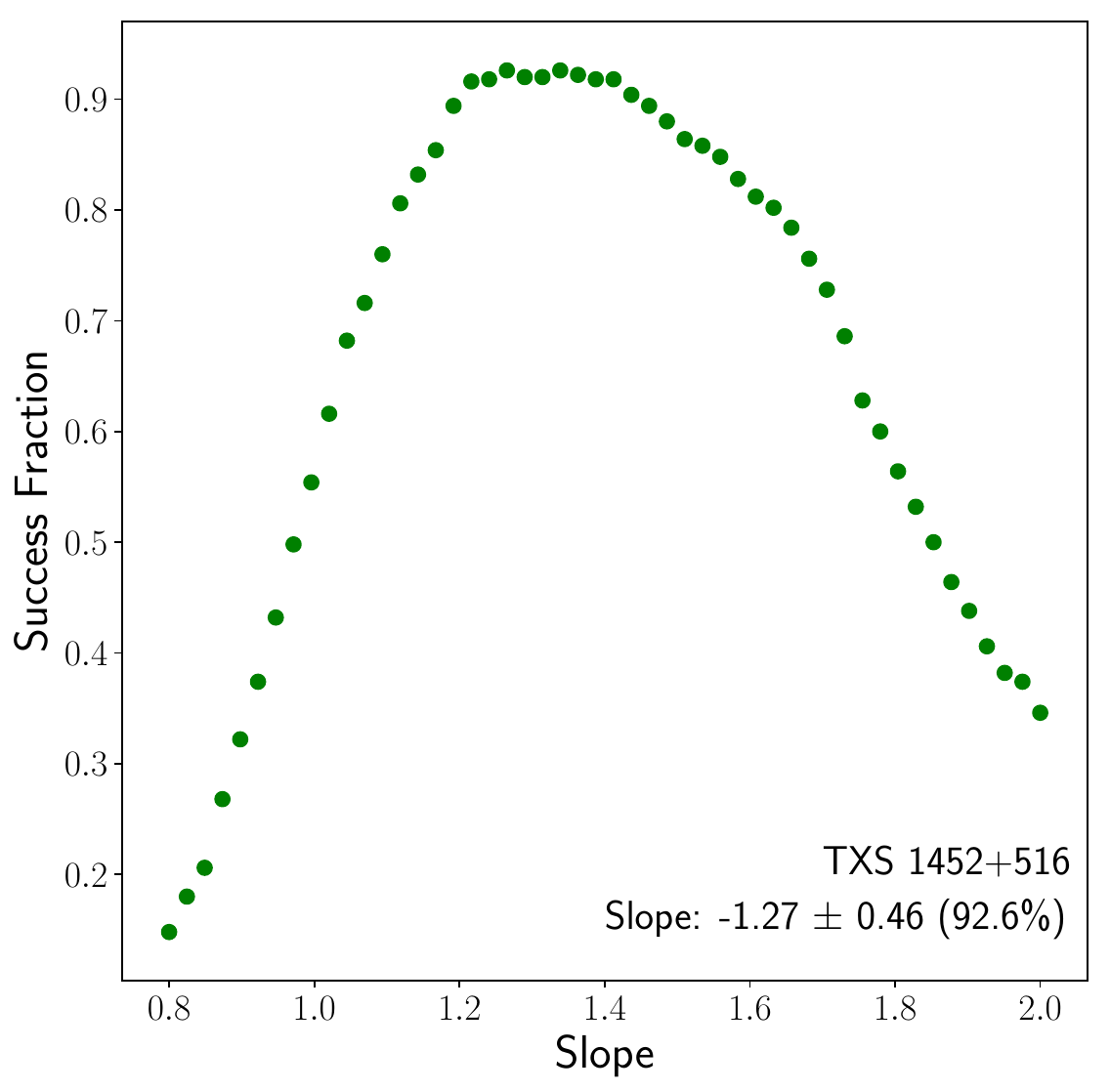}
\caption{Power-law index distribution as obtained using the \texttt{PSRESP} method on the V-band LC of TXS~1425+516.}
\label{PSD_result}
\end{figure}

To obtain the local significance of the correlation and periodicity analysis, we employ the method based on simulating LCs \citep[][]{emma_lc}. These simulated LCs have the same sampling, PSD, and probability distribution function (PDF) as the original LC. Thus, they will have the same statistical properties as the original data set, and accurately modeling the underlying type of noise of the real data.
We calculate the PSD of each LC and obtain the spectral index that describes the derived PSD associated with each data set. The PSD denotes the energy variation as a function of the frequency of the time series and is commonly modeled as a PL function (PSD $\propto f^{-\alpha}$). We estimate the PLs describing the different PSDs using the Power Spectrum Response Method \citep[\texttt{PSRESP}, ][]{uttley2002}. \texttt{PSRESP}\footnote{We use the implementation of \url{https://github.com/wegenmat-privat/psresp}} also provides the uncertainty of the estimated slope and the ``success fraction'' as a measurement of the goodness of the fit. This ``success fraction'' estimates the discrepancy between the data and the fit for each scanned slope value, leading to the slope that best reproduces the derived PSD. To estimate the power spectrum index, we simulate 1,000 LCs using the approach from \citet{timmer_koenig_1995}, with the same observational properties of the original LC, i.e., mean, standard deviation, flux PDF distribution \citep[][]{shah_tk_lognormality}.  \citet{benkhali_power_spectrum} show that both methods of \citet{timmer_koenig_1995} and \citet{emma_lc} yield similar results. We sample the frequency space using a binning of 0.005 and scan slope values between 0.8 and 2.0 using a binning of $\sim$0.02-0.03. An example of a PSD fit using the \texttt{PSRESP} method can be seen in Figure~\ref{PSD_result}. With these PSD slope and PDF estimations, we employ the implementation from \citet{connolly_code} for simulating 20,000 non-periodic LCs.

\subsection{Correlation}
The search for correlations can be performed using either one signal (auto-correlation) or two different signals (cross-correlation). Physically, auto-correlation (cross-correlation) corresponds to a measurement of the similarity of a signal with itself (with respect to a second signal). 

The Discrete Correlation Function (DCF) is the traditional approach to search for correlations \citep{edelson_dcf}. However, as described in \cite{zdfc_alexander}, the DCF presents some inherent problems. First, the DCF adds interpolated points between those from observational data, assuming that the LC varies smoothly, which may be a risky assumption. Another problem of the DCF is the bias in the estimation of the correlation coefficient of each bin, which can produce inconsistencies in the time lag of correlation. To overcome these disadvantages, the \textit{z}-transformed discrete correlation function was implemented \citep[\textit{z}-DCF,~][]{zdfc_alexander}. \textit{z}-DCF avoids the interpolation and performs a correct normalization for the correlation coefficient, ensuring that the absolute value of them is $\leq$1. This technique estimates the correlation function for sparse, unevenly sampled LCs. To perform the correlation, we identify contemporaneous high-flux states across all wavelengths using the Bayesian block algorithm \citep{scargle_bayesian_blocks}. For this work, we compute the cross-correlations of all the MWL data sets with the corresponding \textit{Fermi}-LAT LC for each source. As for the periodicity search methods, the \textit{z}-DCF was also developed for handling unevenly sampled time series with multiple gaps \citep[see ][]{zdfc_alexander}.

No 28-day binning in the MWL LCs (as we do for the periodic analysis) is used for the correlation. The local significance is obtained by applying the same methodology as the previous subsection: simulating 20,000 LCs with the same properties as the original. 

The (auto- or cross-) correlation can also be used to search for periodicities. In such a scenario, the correlation function is expected to have a sinusoidal shape with recurrent and approximately equidistant maxima and minima, separated by a delay corresponding to the signal period. It is also a solid method for periodic, non-sinusoidal series, as it is based on the similarity of the LCs at different time lags, regardless of the shape of the data series. To measure the period, we smooth the correlation function with a Savitzky–Golay filter\footnote{Making use of the function ``savgol filter'' of the Python package ``Scipy''}, which reduces the low-frequency variability without distorting the signal tendency \citep[][]{savitzky_golay_filter}, and identifies the minima and maxima in the resulting output curve. Then, a list of periods can be calculated from the distance between consecutive maxima and minima. The median of the different inferred values is the period of the signal. The uncertainty is obtained by the equation presented by \citet{mcquillan2013} as
\begin{equation}
\sigma\textsubscript{P} = \frac{1.483\times \text{MAD}}{\sqrt{N-1}},
\end{equation} 
where $N$ is the number of peaks in the correlation and $\text{MAD}$ is the median of the absolute deviations of the periods inferred from the different peaks. For this work, we calculate the periodicity using the cross-correlation between all the MWL data with the corresponding \textit{Fermi}-LAT LC for each source and the auto-correlation of such MWL data sets.

\subsection{Variability} \label{sec:variability}
To quantify the variability of the studied blazars, we evaluate their fractional variability ($F_{\text{var}}$) in each wavelength. This parameter was estimated as
\begin{equation}
F_{var}=\sqrt{\frac{S^2-\langle\sigma^2_{err}\rangle}{\langle x \rangle^2}},
\label{fractional_variability_equation}
\end{equation}
following the prescription from \citet{vaughan2003}, where its uncertainty can be expressed as
\begin{equation}
\centering
\sigma_{F_{var}}=\sqrt{ F_{var}^{2}+\sqrt{ \frac{2}{N}\frac{\left \langle \sigma _{err}^{2} \right \rangle^{2}}{\left \langle x \right \rangle^{4}}+  \frac{4}{N}\frac{\left \langle \sigma _{err}^{2} \right \rangle }{\left \langle x \right \rangle^{2}} F _{var}^{2}}} - F_{var},
\end{equation} 
where $S^{2}$ is the variance of the LC, $\langle x \rangle$ is the mean value of the flux in the LC, and $\langle\sigma^2_{err}\rangle$ the mean square error. $F_{\text{var}}$ is affected by the time coverage, sampling, and binning of the data \citep[see ][]{vaughan2003, schleicher2019}. Consequently, we employ a similar time window and binning to analyze MWL data. A detailed discussion on the estimation of the $F_{\text{var}}$ and caveats such as gaps or uneven sampling can be found in \citet{schleicher2019}.

Additionally, we evaluate the timescales of the observed variability using the structure function (SF). The SF estimates the differences between the squared magnitude as a function of the time separation between measurements as
\begin{equation}
\centering
SF(\tau)=[x(t)- x(t+\tau)]^{2},
\end{equation}
where $x(t)$ and $x(t+\tau)$ are measurements separated by a time interval $\tau$. Characteristic variability timescales are reflected as flattenings and dips in the SF \citep[see, e.g., ][]{raiteri2021}. Additionally, the SF can complement the periodicity analysis, as it is expected that periodic variability signatures also appear in the SF as dips at a time interval compatible with the value of the period \citep{wang2017}. While this method cannot be interpreted as an independent test of periodic variations, it should report the presence of that variability signature, indicating that the reported periodic variability timescale is a real timescale present in the data rather than a red-noise artifact. However, \citet{emma_sf} presents the problems of the SF, suggesting that the method could provide incorrect timescales depending on the length of the data set and the shape of the associated PSD. Despite this caveat, this tool can complement the periodicity analysis, reveal other variability timescales, and give some information about the origin of the variability (see Section \ref{sec:discussion}).

\section{Results}\label{sec:results}
\subsection{Periodicity and correlation}
\subsubsection{Periodicity} \label{sec:periodicty_results}
The periodicity analysis is performed on a total of 9 sources with enough temporal coverage on the MWL LCs to detect periods on the order of a few years (more than 2 cycles using the derived period from P22). The results of Table \ref{tab:periodicity_results} also reveal compatible periods with those reported in P20 and P22 in 5 of them: PG 1246+586, S4 1144+40, S4 0814+42, PKS 0301$-$243, and TXS 1902+556. The results of TXS~0518+211 show a slightly longer ($\sim$4~years) period than that reported in P20 and P22 ($3.1 \pm 0.4$~years). We note that the MWL LCs of this source have a shorter coverage w.r.t. the \textit{Fermi}-LAT curve. 
Nevertheless, due to the large uncertainties derived by the periodicity analysis for the different inferred values of the period, the two estimates are still compatible.

The period of 2.8 yr\footnote{All quantities referred to in the text are given without the associated uncertainties to make the text more readable. However, all reported numbers can be found in their respective tables, with the associated uncertainties. We refer the reader to those for details.} obtained for PKS 0447$-$439 is inconsistent with the 1.9-year period from P20 and P22. Similarly, the period of 1.8 yr for MG1~J021114+1051 is inconsistent with the 2.9-year period from P20 and P22. Finally, a period of 2.6 yr is reported for PKS 0208$-$512, which is compatible with P20 (2.7 yr) but inconsistent with P22 (3.8 yr).

The local significance of the results typically ranges from 2$\sigma$ to 5$\sigma$, depending on the method. The shorter time coverage and uneven sampling of the MWL data sets also lead to much larger errors of the derived periods than those from the \textit{Fermi}-LAT LCs. The periodicity-search methods respond differently to the gaps of the LCs (see P20). In fact, as stated in Section~\ref{sec:methodology}, the LSP, GLSP, WWZ and \textit{z}-DCF have been specifically developed for having an improved performance under these conditions \citep[see e.g. ][ for these methods]{lomb_1976,scargle_1982,lomb_gen,wwz,zdfc_alexander}. As demonstrated in P20 and \cite{otero2023}, LSP and WWZ are the most robust against the missing data, and DFT is the most affected method. 

The impact of red noise can be seen in Table \ref{tab:periodicity_results}. Specifically, each method has also a different bias to red noise. The most robust ones are again the GLSP and WWZ, as seen in performance tests developed by P22 and \cite{otero2023}, and the most sensitive is again the DFT. This red noise can also result in false peaks and small shifts \citep[see ][]{otero2023}. For this reason, some methods tend to have incompatible periodicities and local significance (e.g., MG1 J021114+1051 in the radio band). Additionally, when no periodicity is obtained in the original LC, the methods tend to report periods $>$4 years due to red noise. This is because red noise tends to overproduce periodograms with peaks at long periods ($>$4 years), which can distort the estimation of this long-period significance.

To be statistically correct, we consider the look-elsewhere effect \citep{{gross_vitells_trial}}. We infer the global significance by applying the trial factor to the local significance of each periodicity. This correction is approximated by
\begin{equation}\label{eq:trial}
p_{\mathrm{global}}=1-(1-p_{\mathrm{local}})^{N},
\end{equation}
where $N$ is the trial factor. The trial factor results from the combination of the number of independent frequencies we search for periodicity in this work and the number of blazars in our sample (351, see P20). 
In P22, we estimated the number of independent frequencies according to the characteristics of the LCs analyzed (e.g., binning, telescope time) and period range of detection ([1-6] years). To perform this estimation, we consider analyzed LCs that have the same temporal coverage ($\approx$12 years). We also use the same binning for the LCs (28 days). The result is 35 independent frequencies (see P22). Therefore, the trial factor of 351$\times$35. Consequently, the global significance of the periodicity analyses is:

\begin{enumerate}
\item $\approx$2.8$\sigma$ for local significance of $\approx$5$\sigma$ (13\% of the results in Table \ref{tab:periodicity_results} and Table \ref{tab:croscorrelation_results} have a local significance of $\approx$5$\sigma$)
\item $\approx$1.8$\sigma$ for a local significance of $\approx$4.5$\sigma$ (11\% of the results in Table \ref{tab:periodicity_results} and Table \ref{tab:croscorrelation_results} have a local significance of $\approx$4.5$\sigma$)
\item $<$1$\sigma$ for local significance $<$4.5$\sigma$ (76\% of the results in Table \ref{tab:periodicity_results} and Table \ref{tab:croscorrelation_results} have a local significance $<$4.5$\sigma$)
\end{enumerate}

\begin{table*}
\begin{center}
\caption{List of periods and the uncertainties (top) with their associated local significance (bottom) for the periodic-emission candidates according to the organization presented in Table~\ref{tab:candidates_list}. The MCMC sine fitting values only show the uncertainties. The $\gamma$-ray period is obtained from the average period (in years) and uncertainty resulting in P22 (as the average of the periodicity analysis). All periods are in years. Note that some sources have two significant periods (organized by the amplitude of the peak), denoted by $\star$. UV1, UV2, UV3 corresponds to UVOT 1928~\AA, UVOT 2246~\AA~and UVOT 2600~\AA, respectively. $St$ corresponds to the data from the Steward Observatory and 15~GHz to the radio band of OVRO.}
 \label{tab:periodicity_results}
\begin{tabular}{cccccccccc} \hline
(1) & (2) & (3) & (4) & (5) & (6) & (7) & (8) & (9) & (10)  \\ 
\multirow{2}{*}{Name} & $\gamma$-ray period  & \multirow{2}{*}{Band} & Power law  & GLSP & LSP  & WWZ & PDM & DFT-Welch & MCMC Sine \\ 
 & [yrs]  &  & [yrs] & [yrs] & [yrs] & [yrs] & [yrs] &  [yrs] &  [yrs] \\ \hline
 \multirow{2}{*}{TXS 0518+211$\star$} & \multirow{2}{*}{3.1$\pm$0.4} & \multirow{2}{*}{R} & \multirow{2}{*}{$4^{\pm1.3}_{1.4\sigma}$} & $1.4^{\pm0.2}_{5.3\sigma}$ & \multirow{2}{*}{$4.0^{\pm1.1}_{4.4\sigma}$} & \multirow{2}{*}{$4.0^{\pm1.2}_{1.4\sigma}$} & \multirow{2}{*}{$3.7^{\pm0.7}_{2.1\sigma}$} & \multirow{2}{*}{$5.0^{\pm0.6}_{2.2\sigma}$} &  \multirow{2}{*}{$4.2^{+0.7}_{-0.2}$} \\
 &  &  &  & $4.2^{\pm1.2}_{4.6\sigma}$ &  &  &  &  &   \\[0.12cm] \hline
 PKS 0447$-$439 & 1.9$\pm$0.2 & V & $2.8^{\pm0.3}_{2.7\sigma}$ & $2.8^{\pm0.3}_{2.5\sigma}$ & $2.7^{\pm0.2}_{2.7\sigma}$ & $2.9^{\pm0.7}_{3.9\sigma}$ & $2.8^{\pm0.3}_{2.9\sigma}$ & $2.3^{\pm0.7}_{2.1\sigma}$ & $2.7^{+0.1}$\T \\[0.12cm] \hline 
S4 1144+40 & 3.3$\pm$0.6 & 15 GHz & \T $3.3^{\pm0.9}_{1.4\sigma}$ & $3.3^{\pm0.8}_{5.2\sigma}$ & $3.3^{\pm0.7}_{5.1\sigma}$ &  $3.3^{\pm0.8}_{2.2\sigma}$ & $3.3^{\pm0.4}_{4.7\sigma}$ & $5^{\pm1}_{5.0\sigma}$ & $2.9^{\pm0.2}$ \\[0.12cm]	\hline

 PKS 0301$-$243 & 2.1$\pm$0.2 & R & \T $2.4^{\pm0.5}_{2.5\sigma}$ & $2.5^{\pm0.5}_{5.3\sigma}$ & $2.5^{\pm0.5}_{5.1\sigma}$ &  $2.2^{\pm0.6}_{4.6\sigma}$ & $2^{\pm0.6}_{2.5\sigma}$ & $2.3^{\pm0.6}_{3.3\sigma}$ & $2.7^{\pm0.1}$\\[0.12cm]	\hline
\multirow{2}{*}{PG 1246+586} & \multirow{2}{*}{2.2$\pm$0.2} & R &\T $2.5^{\pm0.4}_{2.3\sigma}$ & $2.5^{\pm0.4}_{5.0\sigma}$ & $2.5^{\pm0.4}_{4.8\sigma}$ & $2.3^{\pm0.6}_{4.1\sigma}$ & $2.5^{\pm0.3}_{3.9\sigma}$ & $2^{\pm0.6}_{5.0\sigma}$ & $2.5^{+0.9}_{-0.1}$ \\
 &  & 15 GHz & \T $3.1^{\pm0.9}_{1.9\sigma}$ & $3.1^{\pm0.6}_{5.2\sigma}$ & $3.1^{\pm0.6}_{5.1\sigma}$ & $3.0^{\pm0.9}_{2.2\sigma}$ & $3.2^{\pm0.3}_{4.9\sigma}$ & $3.5^{\pm0.3}_{4.0\sigma}$ & $2.9^{+0.1}_{-2.1}$ \\[0.12cm] \hline
\multirow{2}{*}{TXS 1902+556} & \multirow{2}{*}{3.3$\pm$0.3} & R &\T $3.7^{\pm0.8}_{1.3\sigma}$ & $3.5^{\pm0.8}_{4.3\sigma}$ & $3.7^{\pm0.8}_{5.2\sigma}$ & $3.7^{\pm0.8}_{1.2\sigma}$ & $3.3^{\pm0.9}_{4.8\sigma}$ & $2.1^{\pm0.4}_{1.1\sigma}$ & $3.6^{\pm0.1}$ \\
 &  & 15 GHz &\T $3.7^{\pm0.6}_{1.2\sigma}$ & $3.6^{\pm0.6}_{1.3\sigma}$ & $3.7^{\pm0.8}_{2.3\sigma}$ & $3.7^{\pm0.7}_{1.1\sigma}$ & $3.6^{\pm0.6}_{5.0\sigma}$ & $2.3^{\pm0.4}_{2.0\sigma}$ &  $3.6^{+0.3}_{-0.1}$ \\[0.12cm] \hline
\multirow{4}{*}{S4 0814$+$42$\star$} & \multirow{4}{*}{2.2$\pm$0.2} &\multirow{2}{*}{V} & \T $2.3^{\pm0.6}_{2.6\sigma}$ & $2.3^{\pm0.5}_{4.3\sigma}$ & $2.3^{\pm0.6}_{2.7\sigma}$ & \multirow{2}{*}{$2.2^{\pm0.2}_{2.4\sigma}$} & \multirow{2}{*}{$2.6^{\pm0.3}_{5.0\sigma}$} & \multirow{2}{*}{$2.3^{\pm0.8}_{4.0\sigma}$} &  \multirow{2}{*}{$2.6^{\pm0.1}$} \\
 &  &  &\T $0.9^{\pm0.1}_{4.5\sigma}$ & $0.9^{\pm0.1}_{3.4\sigma}$ & $0.9^{\pm0.1}_{5.3\sigma}$ &  &  &  &   \\
 &  & R &\T $1.2^{\pm0.1}_{3.7\sigma}$ & $1.2^{\pm0.2}_{3.9\sigma}$ & $1.2^{\pm0.1}_{4.2\sigma}$ & $4.5^{\pm1.2}_{2.3\sigma}$ & $3.6^{\pm0.6}_{4.8\sigma}$ & $4.5^{\pm1.3}_{4.2\sigma}$ & $0.8^{\pm0.1}$  \\
 &  & 15 GHz &\T $2.3^{\pm0.6}_{2.4\sigma}$ & $2.3^{\pm0.6}_{4.3\sigma}$ & $2.3^{\pm0.6}_{2.3\sigma}$ & $5.0^{\pm1.3}_{4.4\sigma}$ & $2.3^{\pm0.2}_{5.0\sigma}$ & $3.7^{\pm0.7}_{2.1\sigma}$ &  $0.9^{+1.2}_{-0.1}$  \\[0.12cm] \hline
 \multirow{4}{*}{MG1 J021114+1051} & \multirow{4}{*}{2.9$\pm$0.3} & St-V &\T $1.8^{\pm0.1}_{2.4\sigma}$ & $4.5^{\pm0.7}_{0.6\sigma}$ & $1.8^{\pm0.2}_{2.3\sigma}$ & $1.7^{\pm0.5}_{1.8\sigma}$ & $3.6^{\pm0.8}_{1.9\sigma}$ & $1.7^{\pm0.3}_{1.9\sigma}$ &  $3.5^{+1}_{-1.6}$ \\
 &  & St-R &\T $1.8^{\pm0.1}_{2.5\sigma}$ & $1.9^{\pm0.2}_{0.5\sigma}$ & $1.8^{\pm0.2}_{2.2\sigma}$ & $4.0^{\pm1}_{1.9\sigma}$ & $1.8^{\pm0.9}_{1.8\sigma}$ & $2.3^{\pm0.4}_{2.4\sigma}$ &  $3.9^{+0.1}_{-0.4}$ \\ 
 &  & R &\T $1.8^{\pm0.3}_{2.5\sigma}$ & -- & $2.0^{\pm0.5}_{2.9\sigma}$ & $3.4^{\pm0.4}_{2.0\sigma}$ & $3.7^{\pm0.8}_{2.2\sigma}$ & $2.8^{\pm0.7}_{2.0\sigma}$ & $3.8^{+0.3}_{-0.4}$  \\
 &  & 15 GHz &\T $3.9^{\pm0.5}_{1.2\sigma}$ & $4.0^{\pm0.6}_{4.3\sigma}$ & $3.9^{\pm0.6}_{4.7\sigma}$ & $4.0^{\pm1.1}_{3.9\sigma}$ & $3.8^{\pm0.7}_{4.0\sigma}$ & $4.5^{\pm1}_{4.1\sigma}$ &  $3.5^{\pm0.1}$ \\[0.12cm] \hline 
\multirow{8}{*}{PKS 0208$-$512} & \multirow{8}{*}{3.8$\pm$0.5}  & X-rays &\T $2.9^{\pm0.2}_{2.0\sigma}$ & $2.9^{\pm0.2}_{3.2\sigma}$ & $2.9^{\pm0.2}_{2.8\sigma}$ & $2.8^{\pm0.2}_{2.6\sigma}$ & $2.9^{\pm0.3}_{3.5\sigma}$ & $3.1^{\pm0.5}_{1.9\sigma}$ &  2.9$\pm$0.1 \\
 &  & UV1 &\T $2.7^{\pm0.3}_{3.0\sigma}$ & $2.7^{\pm0.3}_{2.0\sigma}$ & $2.7^{\pm0.3}_{4.0\sigma}$ & $2.8^{\pm0.2}_{3.0\sigma}$ & $2.7^{\pm0.8}_{1.0\sigma}$ & $2.9^{\pm0.4}_{1.4\sigma}$ &  $3.3^{+0.9}_{-0.6}$ \\ 
 &  & UV2 &\T $2.6^{\pm0.3}_{2.1\sigma}$ & $2.7^{\pm0.3}_{1.0\sigma}$ & $2.6^{\pm0.3}_{3.7\sigma}$ & $2.8^{\pm0.1}_{2.6\sigma}$ & $2.7^{\pm0.8}_{1.0\sigma}$ & $3.0^{\pm0.4}_{1.0\sigma}$ &  $4.0^{+0.4}_{-1.2}$ \\
 &  & UV3 &\T $2.6^{\pm0.3}_{1.8\sigma}$ & $2.7^{\pm0.2}_{2.2\sigma}$ & $2.6^{\pm0.3}_{2.8\sigma}$ & $2.8^{\pm0.1}_{2.5\sigma}$ & $2.7^{\pm0.9}_{1.1\sigma}$ & $2.9^{\pm0.4}_{2.0\sigma}$ &  $4.1^{+0.6}_{-1.4}$ \\ 
 &  & B &\T $2.7^{\pm0.3}_{3.0\sigma}$ & $2.6^{\pm0.2}_{5.3\sigma}$ & $2.6^{\pm0.3}_{4.5\sigma}$ & $2.7^{\pm0.2}_{5.0\sigma}$ & $2.7^{\pm0.4}_{1.3\sigma}$ & $4.8^{\pm0.8}_{4.9\sigma}$ &  2.7$\pm$0.1  \\
 &  & V &\T $2.7^{\pm0.3}_{3.1\sigma}$ & $2.6^{\pm0.3}_{4.4\sigma}$ & $2.6^{\pm0.3}_{3.6\sigma}$ & $2.7^{\pm0.2}_{4.6\sigma}$ & $2.7^{\pm0.4}_{2.3\sigma}$ & $2.8^{\pm0.3}_{5.0\sigma}$ &  2.8$\pm$0.1  \\ 
 &  & R &\T $2.7^{\pm0.3}_{3.0\sigma}$ & $2.6^{\pm0.3}_{5.0\sigma}$ & $2.6^{\pm0.3}_{4.0\sigma}$ & $2.7^{\pm0.2}_{5.1\sigma}$ & $2.7^{\pm0.5}_{2.4\sigma}$ & $4.8^{\pm0.8}_{5.0\sigma}$ &  2.8$\pm$0.3 \\
 &  & J &\T $2.7^{\pm0.3}_{2.8\sigma}$ & $2.6^{\pm0.4}_{2.0\sigma}$ & $2.6^{\pm0.5}_{2.5\sigma}$ & $3.8^{\pm0.5}_{3.0\sigma}$ & $2.7^{\pm1.3}_{1.5\sigma}$ & $1.5^{\pm0.1}_{1.9\sigma}$ &  2.7$\pm$0.1 \\[0.12cm] \hline 
\end{tabular}
\end{center}
\end{table*}

\subsubsection{Correlation} \label{sec:correlation_results}

The results of the $\gamma$-ray-MWL cross-correlation analysis are presented in Table \ref{tab:croscorrelation_results}. Additionally, if the cross-correlation obtained with the \textit{z}-DCF shows a hint/evidence of periodicity, the inferred period is also displayed in this table \citep[i.e., the sinusoidal shape of the derived correlation curve with recurrent and approximately equidistant maxima/minima, see ][]{mcquillan2013}. In our notation, positive lags mean that the $\gamma$-ray LC precedes the optical/IR/radio emission. We find that all optical/IR LCs are positively correlated with the $\gamma$-ray emission with time lags compatible with 0 days. Here, all lags $<\pm$28 days are compatible with 0 lag due to the 28-day binning of the \textit{Fermi}-LAT. We also note that since the optical/IR LCs are expressed in magnitudes, negative values of the correlation function represent a positive correlation between the $\gamma$-ray and optical/IR LCs. This is due to the fact that decreasing magnitudes are translated into increasing fluxes. These correlations show significance ranging from 2$\sigma$ to 5$\sigma$ (pre-trials significance). Our results in the optical and IR bands are compatible with those included in \citet{cor_kait_ovro_bigsample}: PG~1246+586, S4~0814+42, TXS~0518+211, TXS~1902+556, MG1~J021114+1051, and TXS 0059+581. PKS 0208$-$512 has a time lag between the $\gamma$-ray and optical bands compatible with 0 days and a local significance of $>$4.0$\sigma$, in agreement with \citet{cor_chatterjee_PKS_0208_512}.

Moreover, radio emission has typically been detected with a delay of a few hundred days \citep[see, e.g. ][]{cor_kait_ovro_bigsample}. These authors report time lags for the radio-$\gamma$-ray correlation of $\approx$160-240 days for TXS~0059+581, S4~0814+42, TXS~0518+211, and TXS~1902+556, with local significance $<$3$\sigma$. We find compatible results for S4~0814+42, with a time lag of 207.2$\pm$28.1 days. However, we do not have radio data for TXS~0059+581 and TXS~0518+211. We also see a positive correlation for S4~1144+40 with a lag of 51.5$\pm$17.3 days. Furthermore, no clear correlation is found between the radio and $\gamma$-ray emission for PG~1246+586 and MG1~J021114+1051. For the latter, a hint of correlation (3.0$\sigma$ of local significance) appears with a time lag of $-$146.5$\pm$31.9 days, contrary to the typical behavior between these bands.  

\begin{figure*}
\includegraphics[width=\textwidth]{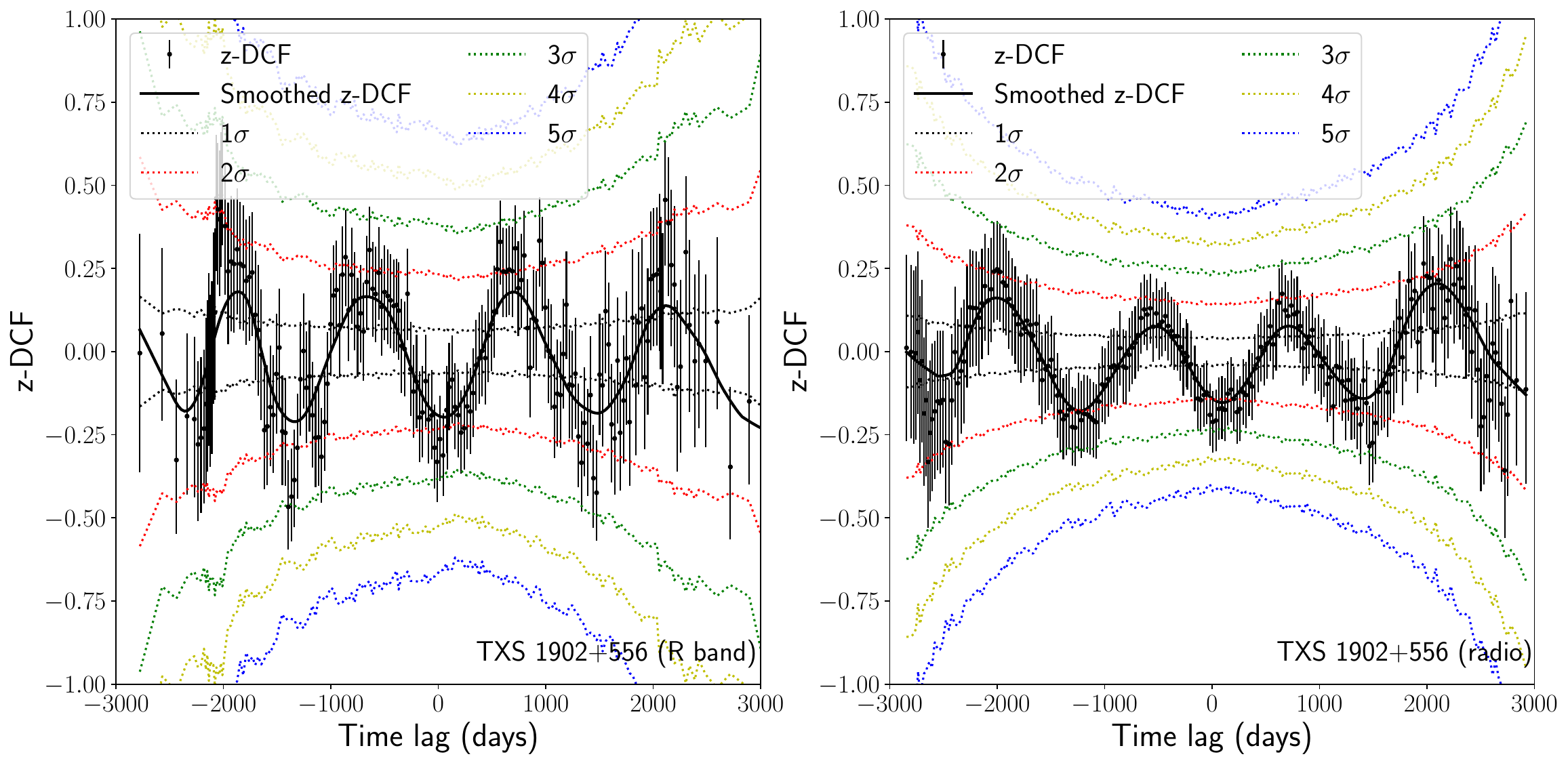}
\caption{\textit{Left:} \textit{Z}-DCF cross-correlation results between the R-band and $\gamma$-ray LCs of TXS~1902+556. \textit{Right:} \textit{Z}-DCF cross-correlation results between the radio and $\gamma$-ray LCs of TXS~1902+556. Both panels display hints of periodic correlation (approximated by the black line which smooths the correlation with Savitzky–Golay filter), denoting that both signals could present similar regular-oscillating behavior with a significance of ~2sigma (pre-trials). The periods are inferred from the distance between the different maxima/minima of the smoothed curves.}
\label{ZDCF_results}
\end{figure*}

Especially interesting is the behavior of TXS~1902+556. 
The cross-correlations between the optical/radio LCs with the $\gamma$-ray emission display a clear sinusoidal behavior expected from a periodic signal \citep[see ][]{mcquillan2013}, with a period of 3.6 $\pm$ 0.2 years and a local significance between 2$\sigma$-3$\sigma$ (see Figure \ref{ZDCF_results}), which is compatible with the results of Table \ref{tab:periodicity_results}. This is also compatible with the results from P22 and supports the hint of periodic behavior for this blazar. The correlation analysis also reveals that the periodic radio emission is delayed half a period w.r.t. the optical and $\gamma$-ray modulations. Through the correlation analysis between the $\gamma$-ray and MWL data sets, other blazars in our sample also show hints of periodicities in their optical LCs, with the recurrent peaks of the sinusoidal z-DCF functions reaching local significances of 2$\sigma$-3$\sigma$. We find compatible periods w.r.t. those from P22 for PKS~0426-380 (V band), PKS~0447-439, PKS~0250-225, S4~1144+40, S4~0814+42, TXS~0518+211, PKS~2052-47 (V and R bands) and 87GB~164812.2+524023. The blazars S4~0814+42, TXS~0518+211 also have compatible periods with Table \ref{tab:periodicity_results}. Regarding PKS 0208$-$512, the cross and the autocorrelations show periods compatible of $\approx$2.7 yr, which again is compatible with P20 but inconsistent with P22. We also observe similar periodicity in X-rays in the autocorrelation, with a lag $\approx$0 days. The periods are reported in Table \ref{tab:croscorrelation_results}. Unfortunately, not all the MWL LCs have enough temporal coverage to display a clear sine-like shape in their cross-correlation.
MG1 J021114+1051 has lags compatible with 0 days in the optical band, but no periodicity is detected with the previous methods (see Table \ref{tab:periodicity_results}). This result is due to the low sampling of the optical LCs.

\begin{table*}
\begin{center}
\caption{List of periods and the uncertainties (top) with their associated local significance (bottom) for the periodic-emission candidates according to the organization presented in Table~\ref{tab:candidates_list}. The MCMC sine fitting values only show the uncertainties. The $\gamma$-ray period is obtained from the average period (in years) and uncertainty resulting in P22 (as the average of the periodicity analysis). All periods are in years. Some sources have two significant periods (organized by the peak amplitude), denoted by $\star$. UV1, UV2, UV3 correspond to UVOT 1928~\AA, UVOT 2246~\AA~and UVOT 2600~\AA, respectively. $St$ corresponds to the data from the Steward Observatory and 15~GHz to the radio band of OVRO. $F_{\text{var}}$ is the fractional variability, and $\tau_{\text{SF}}$ is the main timescale inferred from the SF. The table also shows the period inferred from the cross-correlation with the $\gamma$-ray LC and the auto-correlation. These periods are associated with the significance levels of the peak/valley of the oscillations (see Figure \ref{ZDCF_results}). We report one significance when the significance levels of the peak and the valley are the same.}
 \label{tab:croscorrelation_results}
\resizebox{\textwidth}{!}{%
\begin{tabular}{ccccccccc} \hline
(1) & (2) & (3) & (4) & (5) & (6) & (7) & (8) & (9)  \\ 
\multirow{2}{*}{Name} & $\gamma$-ray period  & \multirow{2}{*}{Band} & \multirow{2}{*}{$F_{\text{var}}$}  & $\tau_{\text{SF}}$ & \multirow{2}{*}{PSD Fit}  & Correlation Lag & Cross-correlation period & Auto-correlation period  \\ 
 & [yrs]  &  &  & [yrs] &  & [days] & [yrs] &  [yrs]  \\ \hline
\multirow{4}{*}{TXS 0518+211} & \multirow{4}{*}{3.1$\pm$0.4} & V &\Tshort 0.35$\pm$0.02 & -- & -- & $-18.3^{\pm22.2}_{5.0\sigma}$ & $3.2^{\pm0.1}_{(2-3)\sigma}$ & -- \\
 &  & St-V &\Tshort 0.32$\pm$0.01 & -- & $1.41^{\pm0.45}_{95.6\%}$ & $9.8^{\pm8.2}_{4.0\sigma}$ & $2.8^{\pm0.3}_{(2-3)\sigma}$ & -- \\
 &  & R &\Tshort 0.36$\pm$0.02 & -- & $1.51^{\pm0.46}_{77.4\%}$ & $18.8^{\pm14.1}_{5.0\sigma}$ & $3.2^{\pm0.1}_{\approx3\sigma}$ & -- \\
 &  & St-R &\Tshort 0.31$\pm$0.01 & -- & -- & $5.5^{\pm12.1}_{3.8\sigma}$ & $3.0^{\pm0.3}_{(2-3)\sigma}$ & -- \\[0.045cm] \hline
PKS 0447$-$439 & 1.9$\pm$0.2 & V &\Tshort 0.34$\pm$0.02 &  2.2$\pm$0.2 &  $1.53^{\pm0.50}_{99.6\%}$ & $2.2^{\pm5.8}_{3.5\sigma}$ & $2.1^{\pm0.2}_{\approx3\sigma}$ & --  \\[0.045cm] \hline
PKS 0250$-$225 & 1.2$\pm$0.1 & V &\Tshort 0.56$\pm$0.02 & -- &  -- & $-11.3^{\pm33.2}_{2.8\sigma}$ & $1.1^{\pm0.1}_{2\sigma}$ & --  \\[0.045cm] \hline
\multirow{2}{*}{S4 1144+40} & \multirow{2}{*}{3.3$\pm$0.5} & V &\Tshort 0.70$\pm$0.01 & -- & $1.24^{\pm0.34}_{96.2\%}$ & $-8.8^{\pm10.4}_{3.0\sigma}$ & $3.5^{\pm0.2}_{(2-3)\sigma}$ & -- \\
 &  & 15 GHz &\Tshort 0.41$\pm$0.01 & 3.6$\pm$0.2 & $1.58^{\pm0.33}_{84.8\%}$ & $51.5^{\pm17.3}_{5.0\sigma}$ & $3.2^{\pm0.2}_{(2-5)\sigma}$ & -- \\ \hline
\multirow{4}{*}{PKS 0301$-$243} & \multirow{4}{*}{2.1$\pm$0.2} & V & 0.70$\pm$0.02 & 2.1$\pm$0.1 & $1.61^{\pm0.43}_{94.8\%}$ & $60.0^{\pm37.7}_{2.1\sigma}$ & -- & -- \\
 &  & R &\Tshort 0.25$\pm$0.01 & 2.2$\pm$0.1 & $1.49^{\pm0.33}_{78.0\%}$ & $21.8^{\pm35.3}_{2.1\sigma}$ & -- & $1.7^{\pm0.3}_{2\sigma}$ \\
 &  & B &\Tshort 0.11$\pm$0.01 & -- & -- & -- & -- & -- \\
 &  & J &\Tshort 0.19$\pm$0.01 & -- & -- & $60.4^{\pm48.9}_{2.0\sigma}$ & -- & -- \\[0.045cm] \hline
 \multirow{4}{*}{PKS 0426$-$380} & \multirow{4}{*}{3.5$\pm$0.5} & V & \Tshort 0.45$\pm$0.01 & 3.2$\pm$0.1 & $1.34^{\pm0.48}_{88.2\%}$ & $3.8^{\pm12.0}_{4.0\sigma}$ & $3.3^{\pm0.2}_{(2-4)\sigma}$ & -- \\
 &  & R &\Tshort 0.43$\pm$0.01 & -- & $1.51^{\pm0.38}_{77.4\%}$ & $13.6^{\pm16.5}_{3.6\sigma}$ & $2.0^{\pm0.4}_{(2-3)\sigma}$ & -- \\
 &  & B &\Tshort 0.48$\pm$0.01 & -- & $1.49^{\pm0.39}_{43.0\%}$ & $13.7^{\pm17.6}_{3.0\sigma}$ & $2.0^{\pm0.5}_{(2-3)\sigma}$ & -- \\
 &  & J &\Tshort 0.38$\pm$0.01 & -- & $1.51^{\pm0.33}_{62.0\%}$ & $3.4^{\pm10.7}_{3.0\sigma}$ & $1.6^{\pm0.3}_{(2-3)\sigma}$ & -- \\[0.045cm] \hline
\multirow{3}{*}{PG 1246+586$\star$} & \multirow{2}{*}{2.2$\pm$0.2} & V &\Tshort 0.31$\pm$0.01 & 2.3$\pm$0.2 & $1.36^{\pm0.37}_{97.4\%}$ & $13.9^{\pm13.1}_{2.4\sigma}$ & -- & -- \\
 & \multirow{2}{*}{1.4$\pm$0.1} & R &\Tshort 0.27$\pm$0.01 & 2.3$\pm$0.2 & $1.46^{\pm0.35}_{89.0\%}$ & $-34.7^{\pm17.4}_{3.0\sigma}$ & -- & -- \\
 &  & 15 GHz &\Tshort 0.71$\pm$0.02 & 1.9$\pm$0.2 & $1.39^{\pm0.24}_{99.6\%}$ & -- & $2.8^{\pm0.1}_{(2-3)\sigma}$ & $1.8^{\pm0.2}_{(1-2)\sigma}$ \\[0.045cm] \hline
\multirow{4}{*}{PKS 2255$-$282$\star$} &  & V &\Tshort 0.56$\pm$0.01 & -- & -- & $22.9^{\pm24.3}_{3.0\sigma}$ & $1.4^{\pm0.1}_{(2-3)\sigma}$ & -- \\
 & 2.8$\pm$0.3 & R &\Tshort 0.45$\pm$0.01 & -- & -- & $9.7^{\pm21.3}_{2.7\sigma}$ & -- & -- \\
 & 1.4$\pm$0.1 & B &\Tshort 0.19$\pm$0.01 & -- & -- & -- & -- & -- \\
 &  & J &\Tshort 0.57$\pm$0.01 & -- & $1.49^{\pm0.58}_{98.6\%}$ & $7.1^{\pm21.8}_{3.0\sigma}$ & -- & -- \\[0.045cm] \hline
\multirow{3}{*}{TXS 1902+556} & \multirow{3}{*}{3.3$\pm$0.3} & V & 0.19$\pm$0.07 & -- & -- & -- &  &  \\
 &  & R &\Tshort 0.23$\pm$0.01 & 3.3$\pm$0.1 & $1.41^{\pm0.25}_{46.0\%}$ & $-11.4^{\pm12.9}_{2.8\sigma}$ &  &  \\
 &  & 15 GHz &\Tshort 0.10$\pm$0.04 & 3.4$\pm$0.1 & $1.39^{\pm0.32}_{37.8\%}$ & $680.5^{\pm28.0}_{2.0\sigma}$ &  &  \\[0.045cm] \hline
\multirow{3}{*}{S4 0814+42} & \multirow{3}{*}{2.2$\pm$0.2} & V &\Tshort 0.28$\pm$0.01 & 2.1$\pm$0.1 & -- & $10.7^{\pm7.0}_{3.0\sigma}$ & $2.4^{\pm0.6}_{(2-3)\sigma}$ & -- \\
 &  & R &\Tshort 0.40$\pm$0.01 & -- & $1.22^{\pm0.20}_{72.6\%}$ & $-15.5^{\pm15.7}_{3.0\sigma}$ & $2.2^{\pm0.2}_{(2-3)\sigma}$ & -- \\
 &  & 15 GHz &\Tshort 0.34$\pm$0.01 & 1.0$\pm$0.1 & $1.61^{\pm0.32}_{77.4\%}$ & $207.2^{\pm28.1}_{4.7\sigma}$ & $2.0^{\pm0.2}_{(1-5)\sigma}$ & -- \\[0.045cm] \hline
 \multirow{5}{*}{MG1 J021114+1051} & \multirow{5}{*}{2.9$\pm$0.3} & V &\Tshort 0.36$\pm$0.03 & -- & -- & $19.1^{\pm25.6}_{2.7\sigma}$ & -- &  \\
 &  & St-V &\Tshort 0.42$\pm$0.01 & -- & $1.29^{\pm0.38}_{83.8\%}$ & $0.8^{\pm11.5}_{3.0\sigma}$ & -- &  \\
 &  & R &\Tshort 0.38$\pm$0.01 & 2.9$\pm$0.1 & $1.12^{\pm0.31}_{52.0\%}$ & $0.8^{\pm11.5}_{2.6\sigma}$ & -- & \\
 &  & St-R &\Tshort 0.42$\pm$0.01 & -- & $1.31^{\pm0.34}_{57.6\%}$ & $0.1^{\pm5.7}_{3.0\sigma}$ & -- &  \\
 &  & 15 GHz &\Tshort 0.42$\pm$0.01 & 2.9$\pm$0.1 & $1.78^{\pm0.40}_{89.4\%}$ & $-146.5^{\pm31.9}_{3.0\sigma}$ & -- &  \\[0.045cm] \hline
\multirow{4}{*}{S3 0458$-$02} & \multirow{4}{*}{3.8$\pm$0.6} & V & 0.62$\pm$0.01 & -- & $1.27^{\pm0.38}_{95.2\%}$ & $8.9^{\pm9.6}_{2.5\sigma}$ &  &  \\
 &  & R &\Tshort 0.35$\pm$0.01 & -- & -- & -- &  &  \\
 &  & B &\Tshort 0.29$\pm$0.01 & -- & -- & -- &  &  \\
 &  & J &\Tshort 0.43$\pm$0.01 & -- & -- & -- &  &  \\[0.045cm] \hline
 \multirow{5}{*}{PKS 2052$-$47$\star$} &  & V &\Tshort 0.84$\pm$0.02 & 1.5$\pm$0.1 & $1.44^{\pm0.24}_{52.4\%}$ & $-8.9^{\pm17.9}_{5.1\sigma}$ & $1.7^{\pm0.2}_{\approx3\sigma}$ & -- \\
 & \multirow{2}{*}{3.1$\pm$0.3} & R &\Tshort 0.77$\pm$0.01 & 1.5$\pm$0.1 & $1.39^{\pm0.36}_{87.4\%}$ & $-10.4^{\pm15}_{4.9\sigma}$ & $1.8^{\pm0.2}_{(2-3)\sigma}$ & -- \\
 & \multirow{2}{*}{1.7$\pm$0.2} & B &\Tshort 0.77$\pm$0.01 & 1.5$\pm$0.2 & -- & $-9.9^{\pm14.1}_{4.7\sigma}$ & -- & -- \\
 &  & J &\Tshort 0.85$\pm$0.01 & 1.5$\pm$0.2 & -- & $-22.6^{\pm8.7}_{4.1\sigma}$ & -- & -- \\
 &  & K &\Tshort 0.75$\pm$0.01 & -- & -- & $-28.6^{\pm17.9}_{3.5\sigma}$ & -- & -- \\[0.045cm] \hline
 MG2 J130304+2434 & 2.1$\pm$0.2 & V &\Tshort 0.81$\pm$0.01 & -- & -- & $11.5^{\pm24.2}_{1.8\sigma}$ & -- & $1.2^{\pm0.1}_{1\sigma}$  \\[0.045cm] \hline
 TXS 0059+581 & 4.0$\pm$0.6 & R &\Tshort 1.06$\pm$0.01 & 2.0$\pm$0.1 &  $1.49^{\pm0.35}_{73.6\%}$ & $-9.8^{\pm14.6}_{5.0\sigma}$ & $2.5^{\pm0.2}_{\approx2\sigma}$ & --  \\[0.045cm] \hline
 TXS 1452+516 & 2.1$\pm$0.3 & V &\Tshort 0.53$\pm$0.01 & 1.8$\pm$0.1 &  $1.27^{\pm0.46}_{92.6\%}$ & $-8.9^{\pm30.1}_{3.7\sigma}$ & -- & $1.3^{\pm0.1}_{(2-1)\sigma}$  \\[0.045cm] \hline
\multirow{9}{*}{PKS 0208$-$512} & \multirow{9}{*}{3.8$\pm$0.5} & X-rays & 0.26$\pm$0.03 & 0.4$\pm$0.1 & X & $5.8^{\pm0.1}_{>4\sigma}$ & -- & -- \\ 
 &  & UV1 &\Tshort 0.38$\pm$0.02 & 0.4$\pm$0.1 & $0.96^{\pm0.67}_{88.6}$ & $-1.1^{\pm0.1}_{>4\sigma}$ & -- & -- \\
 &  & UV2 &\Tshort 0.32$\pm$0.01 & 0.4$\pm$0.1 & $0.98^{\pm0.68}_{39.2}$ & $-1.1^{\pm0.1}_{\approx4\sigma}$ & -- & -- \\
 &  & UV3 &\Tshort 0.35$\pm$0.01 & 0.5$\pm$0.1 & $1.22^{\pm0.58}_{84.1}$ & $-1.1^{\pm0.1}_{>4\sigma}$ & -- & $2.5^{\pm0.1}_{(2-1)\sigma}$ \\
 &  & B &\Tshort 0.84$\pm$0.01 & 0.9$\pm$0.1 & $1.28^{\pm0.56}_{74.5}$ & $-5.8^{\pm0.2}_{\approx5\sigma}$ & $2.5^{\pm0.1}_{2\sigma}$ & $2.4^{\pm0.1}_{(2-1)\sigma}$ \\
 &  & V &\Tshort 0.81$\pm$0.01 & 0.8$\pm$0.1 & $1.32^{\pm0.46}_{70.7}$ & $4.0^{\pm0.1}_{\approx5\sigma}$ & $2.4^{\pm0.2}_{2\sigma}$ & $2.6^{\pm0.1}_{(2-1)\sigma}$ \\
 &  & R &\Tshort 0.91$\pm$0.01 & 0.7$\pm$0.1 & $1.30^{\pm0.89}_{98.2}$ & $-5.8^{\pm0.1}_{\approx5\sigma}$ & $2.4^{\pm0.2}_{2\sigma}$ & -- \\
 &  & J &\Tshort 0.81$\pm$0.01 & 0.7$\pm$0.1 & $1.43^{\pm0.53}_{80.8}$ & $-6.3^{\pm0.2}_{\approx5\sigma}$ & $2.4^{\pm0.3}_{2\sigma}$ & -- \\ 
 &  & K &\Tshort 0.72$\pm$0.01 & 0.7$\pm$0.1 & $1.18^{\pm0.36}_{94.7}$ & $1.8^{\pm9.1}_{>3\sigma}$ & $2.3^{\pm0.2}_{1\sigma}$ & -- \\[0.045cm] \hline
87GB 164812.2+524023 & 2.8$\pm$0.3 & V &\Tshort 0.42$\pm$0.01 & -- &  -- & $-15.8^{\pm29.4}_{2.1\sigma}$ & $2.5^{\pm0.2}_{\approx2\sigma}$ & --  \\[0.045cm] \hline
\end{tabular}}
\end{center}
\end{table*}

\subsection{Variability} 
\subsubsection{Fractional Variability}
The results of the variability analysis are shown in Table \ref{tab:croscorrelation_results}. For 4 out of 5 sources observed by OVRO, we find that the $F_{\text{var}}$ is smaller or similar to that in optical/IR wavelengths. Lower $F_{\text{var}}$ in the radio band has been the common behavior observed in the past in variability studies. PG~1246+586 is the only source of our sample with a significantly higher $F_{\text{var}}$ in radio, with $F_{\text{var}}^{\text{radio}}=0.71 \pm 0.02$, compared to $F_{\text{var}}^{\text{V, R}}=0.27 \pm 0.01$ found in the V and R optical bands. This indicates that the emission is more variable in radio than in optical wavelengths for this blazar. However, the coverage and sampling of the radio data set are higher than that in the latter wave bands. By computing the $F_{\text{var}}$ on the quasi-simultaneous radio and optical data, we obtain again the typically observed behavior for blazars, with a lower $F_{\text{var}}=0.14\pm0.02$ in radio. Moreover, the derived values of the $F_{\text{var}}$ in the different optical/IR bands are typically compatible with errors. Significant discrepancies (e.g., B band of PKS~2255-282 or B band of PKS~0301-243) are explained as large differences in time coverage of the LCs as for the case of PG~1246+586 already mentioned. In these cases, the $F_{\text{var}}$ of the simultaneous data sets are found to show compatible values. 

\begin{figure}
\includegraphics[width=\columnwidth]{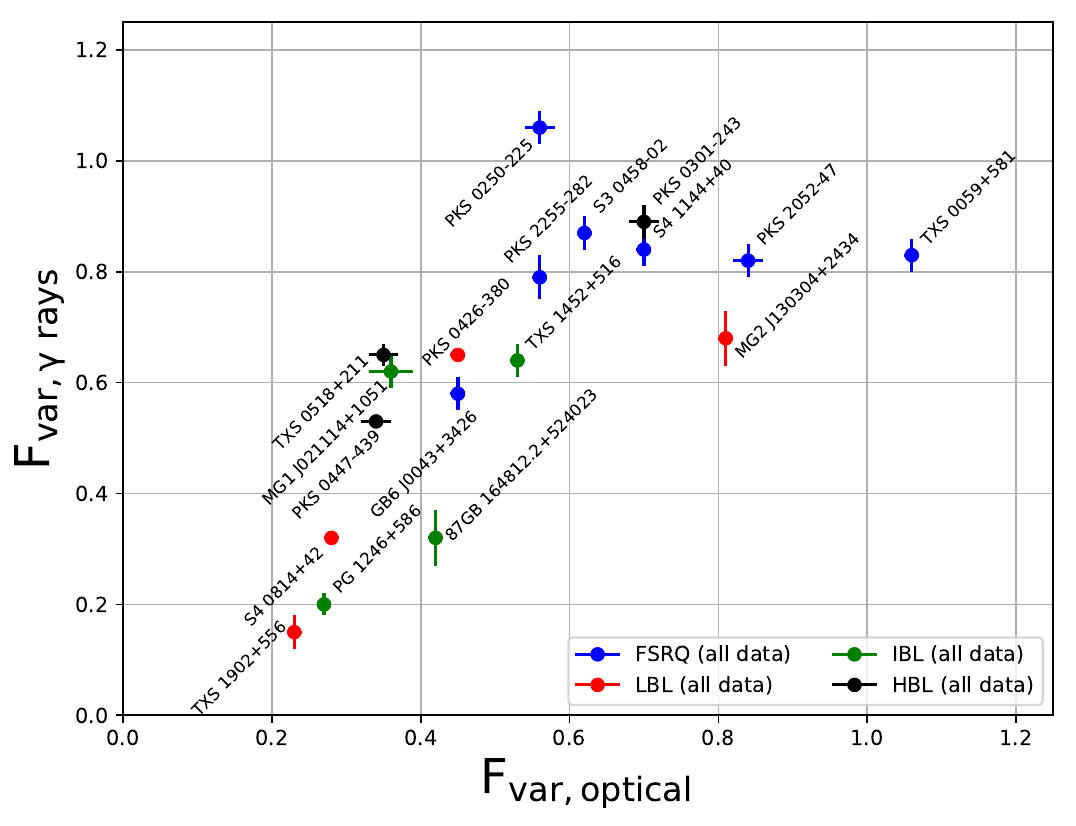}
\includegraphics[width=\columnwidth]{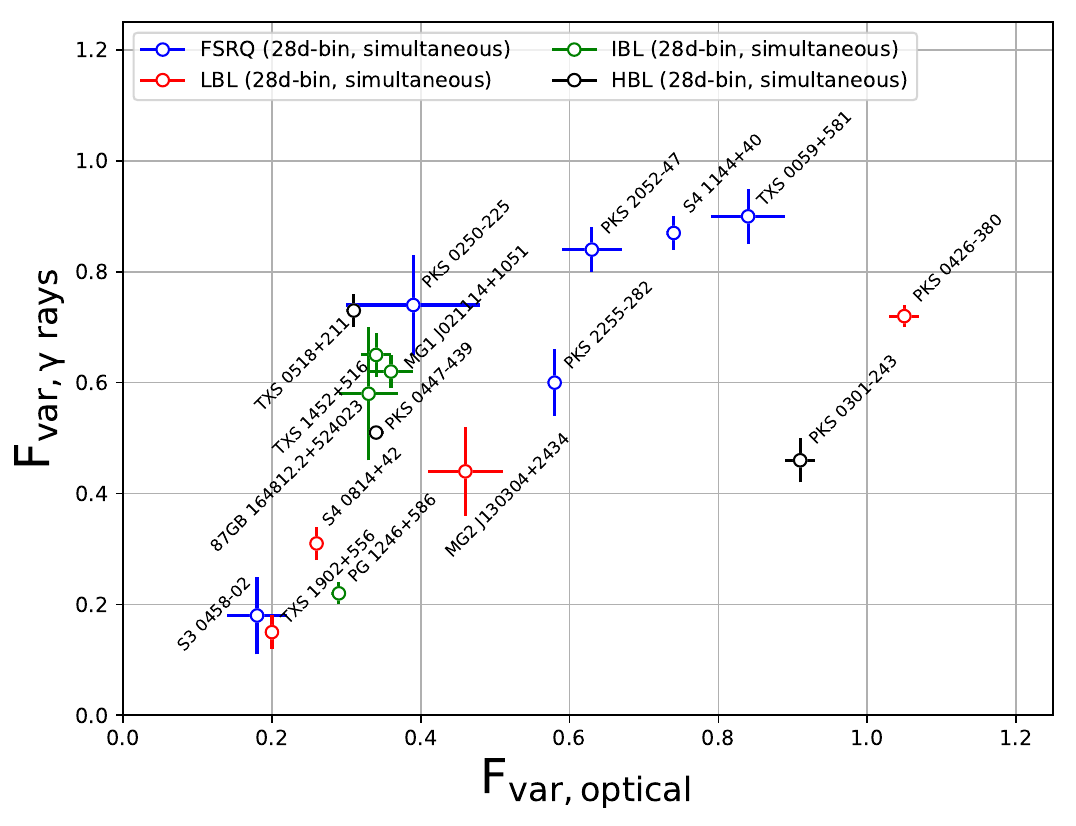}
\caption{Optical fractional variability vs. $\gamma$-ray fractional variability. Blue markers represent the FSRQs of the blazar sample. Red, green, and black markers correspond to the different BL Lac subtypes (LSPs, ISPs, and HSPs, respectively). \textit{Top}: $F_{\text{var}}$ obtained using the complete data sets. \textit{Bottom}: $F_{\text{var}}$ obtained after applying a 28-day binning to the optical data and using the simultaneous \textit{Fermi}-LAT data.}
\label{Fvar_results}
\end{figure}

We also evaluate the differences in the $F_{\text{var}}$ between BL Lac objects and flat spectrum radio quasars (FSRQs). We find that, while BL Lac objects show rather low $F_{\text{var}}$ values (typically between 0.2 to 0.4), FSRQs display a much higher variability ($F_{\text{var}} \sim $ 0.45--0.85), as can be seen in Figure \ref{Fvar_results}, which shows the distribution of the optical $F_{\text{var}}$ with respect to that in the $\gamma$-ray regime for the different blazar types. 

The estimation of the $F_{\text{var}}$ is highly affected by the time coverage, sampling, and binning of the data \citep[see ][]{vaughan2003, schleicher2019}. To test the effect of the different sampling in each wave band, we have also estimated the $F_{\text{var}}$ with all the simultaneous MWL and $\gamma$-ray data with a matching 28-day binning. This estimation is also represented in Figure \ref{Fvar_results} with open markers. We observe the same trend as before, where FSRQs tend to be more variable than BL Lac objects. Also, except for a few objects, the optical $F_{\text{var}}$ derived for the simultaneous data do not change significantly, pointing towards a dominant long-term variability over shorter timescales. 

Finally, as shown in Figure \ref{Fvar_results}, blazars with better sampled MWL LCs and more simultaneous data (e.g., TXS 0518+211) exhibit optical $F_{\text{var}}$ values that are consistent between the complete data set and the binned data set. For these blazars, this indicates that long-time scales dominate the variability. For sources with a difference between the $F_{\text{var}}$ before and after the binning of the data, there may be significant variability in shorter timescales. Nevertheless, these sources typically correspond to those with the poorest sampling. Therefore, more data are needed to evaluate the presence of significant, faster variability. 

\subsubsection{Characteristics Timescales}
Characteristic variability timescales are also evaluated from the SF. We find timescales compatible within uncertainties with the derived periods by P22 and in this work (see Table \ref{tab:croscorrelation_results}). However, some discrepant values between the suggested timescales by the SF and the derived periods are observed (for instance, TXS~0059+581, where the derived times scale corresponds to half of the period reported by P22).

\begin{figure*}
\includegraphics[width=\textwidth]{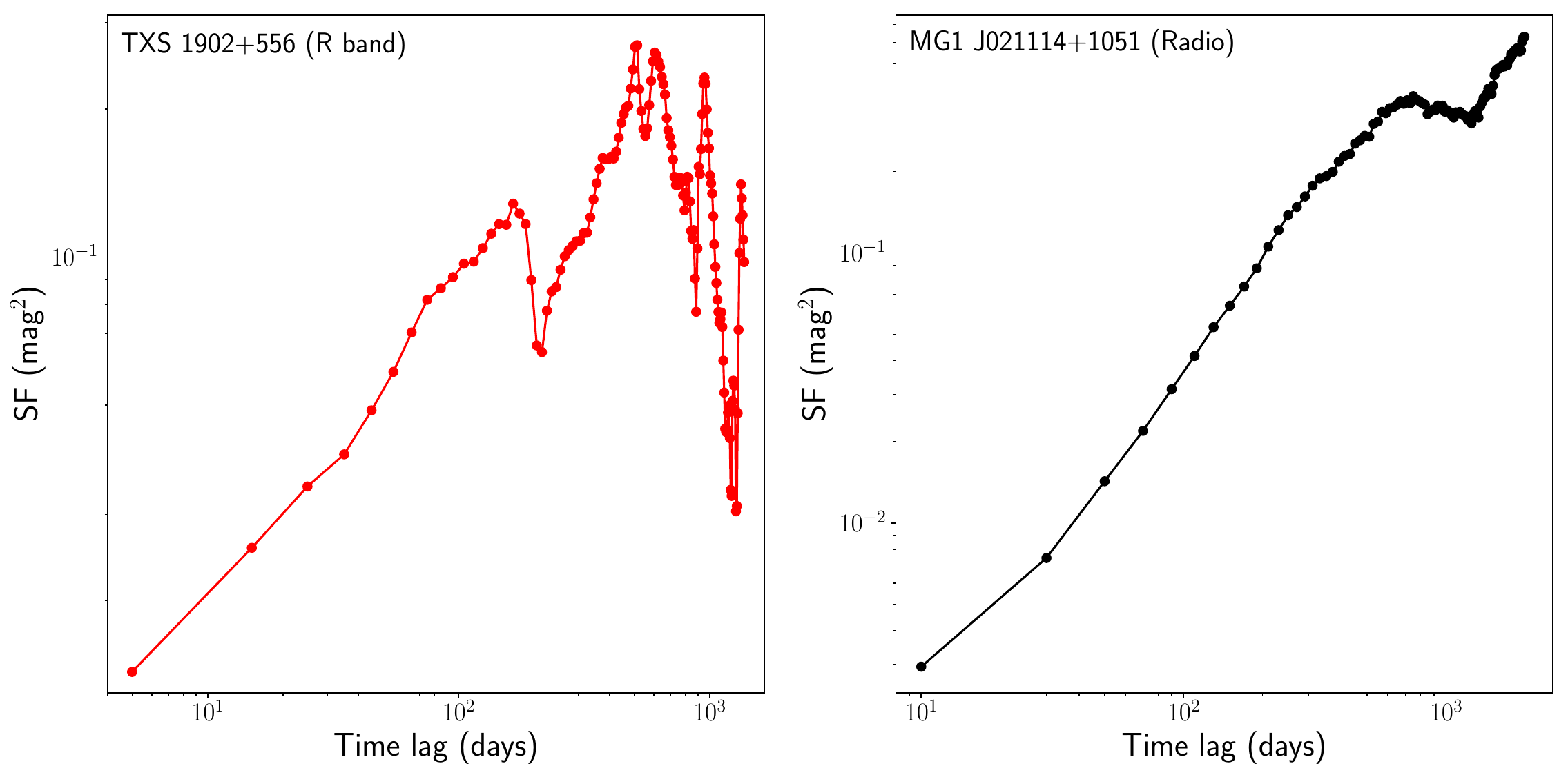}
\caption{Results of the SF analysis. \textit{Left:} TXS~1902+556 (R band). \textit{Right:} MG1 J021114+1051 (Radio).}
\label{SF_results}
\end{figure*}

Additionally, for some sources, we find more characteristic variability timescales. These timescales show faster variability on the order of a few tens or hundred days. In Table \ref{tab:croscorrelation_results}, we only show those compatible with the reported periodicities. An example of two SFs in the optical and radio bands is shown in Figure~\ref{SF_results}.

\subsubsection{Power Spectrum Index and Flux Distribution}
We estimate the PSDs of each target and the best-fit PL functions that describe these PSDs with the \texttt{PSRESP} package. The resulting indices of the PLs are shown in Table \ref{tab:croscorrelation_results}. We find that the PL indices range from 1.12 to 1.78. No significant differences in the PSD slope are detected between the BL Lac (with mean $\alpha$=1.37$\pm$0.27) and FSRQ (with mean $\alpha$=1.43$\pm$0.39) subsamples. Moreover, we see that the mean radio PSDs have slightly steeper slope values (with mean $\alpha$=1.55$\pm$0.32) than the optical/IR slopes (with mean $\alpha$=1.40$\pm$0.35). However, the sample of blazars with radio data is smaller than that of blazars with optical/IR data (4 vs. 12), which can distort this comparison. 

In addition to that, we also evaluate both the flux distributions used for the artificial LC simulation with a normal and a log-normal PDF. We observe that the log-normal distribution provides a better fit to the data. 

\section{Discussion}\label{sec:discussion}
\subsection{Periodicity}
The periodicity analysis performed on the MWL LCs finds hints of periods compatible with those reported for the $\gamma$-ray LCs in P20 and P22 in some of the analyzed blazars: S4 1144+40, S4 0814+42, TXS 0518+211, TXS 1902+556, PG 1246+586 and PKS 0301$-$243 (only in the last two blazars the pre-trials median significance is $\gtrsim$3.5$\sigma$, $\approx$0$\sigma$ post-trials). This significance is obtained as the median of the results of Table \ref{tab:periodicity_results} for each method. Note that this median significance does not have an actual statistical meaning; rather, it is used as an arbitrary way of combining all of the significance for a given source to sort the candidates and compare with the results obtained in P22 for the $\gamma$-rays.

This periodicity has been interpreted in the past within several models. The most common interpretations can be summarized in models based on the existence of a binary SMBH system or models based on geometrical effects. 

Binary SMBH systems are a possible explanation for periodic emissions across the electromagnetic spectrum, as seen in OJ~287 and PG~1553+113 \citep[][, respectively]{agudo2011, ackermann_pg1553}. High-resolution simulations confirm that such systems can have yearly orbital periods, influencing jet luminosity through various factors like binary eccentricity, SMBH mass, mass ratio, and the binary's evolutionary stage \citep[][]{WS_smbbh, zrake_eccentricity}. These factors produce a consistent periodic pattern across different wavebands. Another hypothesis involves a helical jet caused by one black hole affecting the other, as suggested by studies on AO~0235+164 \citep[see ][]{ostorero_jet_2004,raiteri2006}, where the emission's periodic modulation is due to changes in the jet's orientation and the resulting variations in relativistic boosting.

Alternatively, geometrical models are also used to explain MWL periodicity, attributing it to jets influenced by strong magnetic fields. These models include precession-jet \citep[e.g., ][]{villata_jet_1999}, rotation-jet \citep[e.g., ][]{hardee_rotation_jet}, and helical structures \citep[e.g., ][]{ostorero_jet_2004}, with helical jet models effectively explaining periodicities in blazars like PKS 1830$-$211 and PKS 2247$-$131 \citep[][, respectively]{nair2005, zhou2018}. Another possibility suggests a twisted jet with periodic orientation changes, impacting the Doppler factor and viewing angle, similar to the effects seen in SMBH binary systems. For example, OJ~287's periodicity has been linked to variations in the Doppler factor due to jet helicity \citep[][]{butuzova2020}.

\subsection{Correlation}
The results of the correlation analysis between the $\gamma$-ray and optical/IR bands reveal a high degree of correlation with time lags compatible with $<$28 days. This behavior implies a co-spatial origin of both emissions, typically expected from leptonic models \citep{cor_kait_ovro_bigsample} rather than from hadronic models \citep{bottcher_2007}. 

In leptonic models, the optical emission is produced by synchrotron radiation while the $\gamma$-rays are typically produced by inverse Compton (IC) scattering of either synchrotron photons \citep[Synchrotron Self-Compton, SSC; e.g. ][]{maraschi1992, berg_emission_models, rajput_bllacs_correlation} or radiation originating from outside of the jet from either the broad-line region or the dusty torus \citep[External Compton, EC; e.g. ][]{sikora_1994, pacciani_fsrqs}. In BL Lacs, particularly HBLs, the process is expected to be SSC, whereas in FSRQs, it is expected to be EC. These strong correlations between the optical and $\gamma$-ray emissions have been observed in the past for other blazars, favoring the leptonic scenario since the same population of electrons is responsible for both emissions \citep[see, e.g. ][]{liao2014}. On the other hand, in hadronic scenarios, the high- and low-energy contributions are caused by different particle populations that are not necessarily easily related. Specifically, protons (and possibly higher-Z nuclei) are responsible for the high-energy emission, and electrons for the low-energy emission \citep[][]{cerruti_emission_models}.

\citet{cor_kait_bigsample} claim that FSRQs typically have positive lags, i.e., $\gamma$-rays leading the optical since they are dominated by the EC, while no evident prevailing lag is observed in BL Lacs, with lags ranging between $-$40 days and 40 days, approximately. The results reported in Table \ref{tab:croscorrelation_results} find time lags in our sample of blazars consistent with such short time delays. The optical emission is typically highly correlated with the $\gamma$-ray emission with no delay ($<$28 days) for all the sources analyzed here. 
The co-spatial origin can also be applied to the X-rays for PKS~0208$-$512 since the $\gamma$-ray and the X-ray emissions are correlated without lag with a local significance of $>$4$\sigma$.  

On the contrary, radio emission is typically delayed a few hundred days \citep[see, e.g. ][]{cor_kait_ovro_bigsample}. We find compatible results with those from \citet{cor_kait_ovro_bigsample} for 4 of the 5 sources with radio data available. The most common explanation is that radio emission comes from an outer part of the jet but is triggered by the same physical mechanism \citep{max-moerbeck2014}. This is due to the high opacity and self-absorption suffered by radio photons in the inner regions. Thus, radio wavelengths are observable from regions of the jet located further away from the central engine. However, other scenarios, such as differences in the cooling times of the electron populations responsible for the radio and optical emissions, have also been proposed \citep[see for instance ][]{bai2003}. 

\subsection{Fractional Variability}
The fractional variability estimation reveals that FSRQs are typically more variable than BL Lacs. Historically, concerning the BL Lac objects, it has been reported that these sources show lower (higher) variability when the synchrotron peak is located at higher (lower) frequencies \citep[e.g. ][]{otero2022}. The results found here are in line with this trend observed between the fractional variability and the frequency of the synchrotron peak reported in several works \citep[see, e.g., Figure 26 from ][, where the variability is quantified for all the sources included in the second \textit{Fermi}-LAT catalog for AGNs, the 2LAC]{ackermann2011}. Here, we observe that ISPs and HSPs show lower $F_{\text{var}}$ than FSRQs. Despite the higher variability of the latter, the low number of each type of BL Lacs (especially HSPs) does not allow us to see a clear trend of the variability of the different subtypes with the frequency of the synchrotron peak. Nevertheless, the smaller sample analyzed here is consistent with the results reported for the complete 2LAC catalog. The existence of this trend has been reported by other works like, for instance, \cite{rajput2020b,bhatta2020}. \cite{ackermann2011} explain these differences in terms of the different cooling timescales of the electrons between the EC and SSC processes of FSRQs and BL Lacs, respectively. Complementary phenomena such as a higher jet power, more efficient accretion disc, and higher electron energies in FSRQs with respect to BL Lac objects lead to a higher variability for the former \citep[][]{hovatta_fractional_variability}. Finally, this combination of phenomena also explains the differences between BL Lacs associated with the frequency of their synchrotron peak \citep[][]{ackermann2011}.

Furthermore, we see that typically, the $F_{\text{var}}$ derived for the MWL data set is lower or of the same order as that of the $\gamma$-ray \textit{Fermi}-LAT LCs. A deep study on the structure of the fractional variability as a function of the energy can also reveal important information regarding the particle population and/or processes producing the broadband emission \citep[see, e.g. ][]{aleksic2015b}. For instance, Mrk~421 shows a $F_{\text{var}}$ with its maximum in the X-ray regime, with a second bump in $\gamma$-rays. In this case, this has been interpreted as a higher variability of the high-energy electron population responsible for the high-energy part of the synchrotron emission observed for this source at X-ray wavelengths.

Alternatively, a progressively increasing $F_{\text{var}}$ from radio to $\gamma$-rays was observed for Mrk~501 \citep{aleksic2015a}. This could indicate that for this source, the variability comes from a combination of the low- and high-energy electron populations in the Thomson and Klein-Nishina regimes, respectively. In addition, the $F_{\text{var}}$ structure can also change with time, indicating that the processes and particle populations dominating the variability can vary. For example, the double peak structure was also observed for Mrk~501 by \cite{furniss2015}. Therefore, by studying and quantifying the broadband variability of blazars, we can understand the importance of each particle population in the overall variability. However, the lack of ultraviolet and X-ray data for the sources analyzed here does not allow us to draw any reliable conclusion concerning the broadband structure of the $F_{\text{var}}$. More MWL data are needed to extract a firm conclusion in this regard.

\subsection{Structure Function}
The variability analysis performed through the SF reveals the characteristic timescales of the dominant long-term flux variations. For those sources for which the temporal coverage of the data allowed us to perform this analysis, the SF analysis measures variability timescales compatible with the periods previously reported by P20 and P22. Additionally, several shorter timescales are observed in the SFs. The radio is the most stable band with almost no timescales displayed by the SF (see right panel of Figure~\ref{SF_results}). On the other hand, optical/IR wavelengths are characterized by a variety of variability timescales, ranging from a few tens of days to several hundred days (Figure~\ref{SF_results}, left panel).

Variability timescales inferred with the SF have been interpreted within several models of different natures. \citet{kawaguchi_variability} explain these timescales based on instabilities in the accretion disc fluctuating in time. Specifically, these timescales would be associated with avalanches of matter on the accretion disc at the same timescales. Three scenarios are presented in \citet{hawkins2002} to interpret the observed variability. Different timescales are associated with instabilities in the accretion disc in the first models, provoked by the injection of matter with the same timescales. An alternative model associates the variability with supernova events. Finally, in the third scenario, different timescales are induced by microlensing effects caused by MACHOs (dark, compact bodies of planetary sizes), leading to timescales of a few years. The characteristic timescales and slopes of the SFs may indicate which scenario can take place. The average slope of the SFs analyzed in this work is 0.5$\pm$0.1, compatible with the 0.44 $\pm$ 0.03 slope derived by \citet{hawkins2002}, representative of the accretion disc model. These results are in line with the ones reported by \citet{devries2006}, where the lensing and starburst models are disfavoured as the origin of the variability in AGNs, supporting the disc instabilities as the cause of the observed variability. This slope also confirms the long-term nature of the variations. \citet{kataoka2001} claim that short-term variability is modeled in the SF as a PL with index $\sim$2.0. However, the slope flattens and adopts lower values for longer timescales, as is the case in this analysis. Additionally, the shorter timescales revealed by the SF are explained through models based on instabilities, turbulence, or shocks propagating along the jet \citep[see for instance ][]{marscher1985,marscher2014}.
We also compare the mean SF slopes for BL Lacs and FSRQs. No significant differences were found between the slope derived for BL Lacs (0.59$\pm$0.10) and the one obtained for the FSRQs (0.50$\pm$0.10). 

Moreover, we obtain similar values for the optical and radio SF slopes for 3 of the 5 sources with OVRO data: PG~1246+586 ($\alpha_{V}=0.61$, $\alpha_{R}=0.60$ and $\alpha_{radio}=0.63$), S4~0814+42 ($\alpha_{V}=0.47$, $\alpha_{R}=0.53$ and $\alpha_{radio}=0.55$) and TXS~1902+556 ($\alpha_{V}=0.51$ and $\alpha_{radio}=0.56$). According to the results presented by \citet{collier2001}, this could be evidence of a common mechanism producing both emissions. On the other hand, MG1~J021114+1051 shows a higher slope for the radio SF ($\alpha_{radio}=1.26$) w.r.t. the V- and R-band SFs ($\alpha_{V}=0.42$ and $\alpha_{R}=0.41$). This difference between optical and radio can explain the different periods for these wavebands shown in Table \ref{tab:periodicity_results} (1.8 and 4 yrs, respectively). The low temporal coverage of the V-band LC of S4~1144+40 does not allow us to compare its slope with the one obtained from the radio data set.

\subsection{PSD characterization}
We also evaluate the PSDs of the different LCs. These PSDs are fitted to a PL function to study the type of underlying noise associated with their variability. The range of spectral indices obtained here is placed between the flickering (pink) and red noise regimes, as expected from blazars with rapid flares and short timescale fluctuations, and long-term (periodic) variability \citep[see ][]{abdo2010}. A pure pink noise signal is described as a PL with a dependency of $1/f$ with the frequency \citep{hufnagel1992}. On the other hand, pure red noise processes are described as $1/f^{2}$ and are typically associated with the erratic and stochastic variability displayed by blazars \citep{vaughan2005}.
Periodic and stochastic variability in different timescales can co-exist in these sources, leading to the admixture of pink and red noise nature, with indices ranging between 1.0 and 2.0. This is supported by SFs that reveal variability on different timescales that are not necessarily associated with the periodic modulation of the emission and can be stochastic in nature, which will lead to the observed spectral indices.

\subsection{Flux Distribution}
A log-normal distribution fits the flux distributions analyzed. This result reflects the influence of the accretion disc on the variability. Indeed, fluctuations in the accretion disc at different radii are propagated inwards to produce an aggregate multiplicative effect in the innermost disc. This disturbance is then transmitted to the jet, generating the log-normal distribution of the variability observed in the MWL emission \citep{shah_lognormality}. The usual normal distribution scenario is associated with additive processes, where independent events are linearly added \citep[see, e.g. ][ and references therein]{biteau_minijets}. However, a log-normal distribution does not necessarily imply multiplicative processes \citep[][]{scargle_lognormal_additive}. 

\section{Summary}\label{sec:conclusions}
We analyze the $\gamma$-ray and MWL (optical, IR, and radio) data of 19 blazars of P22, showing hints of periodic modulation in 5 of them with $\geq$3.0$\sigma$ ($\approx$0$\sigma$ post-trials). The $\approx$0$\sigma$ in the post-trials significance does not allow to claim any presence of periodicity in these sources. However, observing the same period could indicate similar regular-oscillating behavior in the MWL emission of such blazars. Therefore, these 5 blazars are promising candidates to be monitored in the next years.     

Moreover, the correlation analysis reveals a high correlation between the $\gamma$-ray, optical, and IR bands with delays $<28$ days and the radio band with typical delays of a few hundred days. This can be the result of the radio being emitted from an outer region of the jet due to the absorption suffered at these wavelengths in the inner regions. Additionally, the variability analysis performed with the SF shows variability timescales compatible with the periods associated with the emission. This result supports the existence of a periodic modulation in the emission of 10 blazars. Shorter timescales revealed by the SF support that such variability could originate from instabilities or fluctuations in the accretion disc.

A combination of pink and red noise for the emission of these objects is confirmed through the PSD analysis. The spectral indices derived from this fit range from 1.12 to 1.78. Finally, the flux distribution analysis shows that a log-normal distribution successfully fits the PDF of the MWL LCs of these objects. These PDFs are characteristic of multiplicative variability processes due to fluctuations in the accretion disc transmitted to the jet.

\section*{Acknowledgements}

P.P and M.A acknowledge funding under NASA contract 80NSSC20K1562. 
S.B. acknowledges financial support by the European Research Council for the ERC Starting grant MessMapp, under contract no. 949555, and by the German Science Foundation DFG, research grant “Relativistic Jets in Active Galaxies” (FOR 5195, grant No. 443220636).

J.O.S. thanks the support from grant FPI-SO from the Spanish Ministry of Economy and Competitiveness (MINECO) (research project SEV-2015-0548-17-3 and predoctoral contract BES-2017-082171).

J.O.S., J.A.P. and J.B.G. acknowledge financial support from the Spanish Ministry of Science and Innovation (MICINN) through the Spanish State Research Agency, under Severo Ochoa Program 2020-2023 (CEX2019-000920-S).

J.O.S. acknowledges financial support from the Severo Ochoa grant CEX2021-001131-S funded by MCIN/AEI/ 10.13039/501100011033. 

A.D. is thankful for the support of the Ram{\'o}n y Cajal program from the Spanish MINECO, Proyecto PID2021-126536OA-I00 funded by MCIN / AEI / 10.13039/501100011033, and Proyecto PR44/21‐29915 funded by the Santander Bank and Universidad Complutense de Madrid.

The \textit{Fermi} LAT Collaboration acknowledges generous ongoing support from a number of agencies and institutes that have supported both the development and the operation of the LAT as well as scientific data analysis. These include the National Aeronautics and Space Administration and the Department of Energy in the United States, the Commissariat \`a l'Energie Atomique and the Centre National de la Recherche Scientifique / Institut National de Physique Nucl\'eaire et de Physique des Particules in France, the Agenzia Spaziale Italiana and the Istituto Nazionale di Fisica Nucleare in Italy, the Ministry of Education, Culture, Sports, Science and Technology (MEXT), High Energy Accelerator Research Organization (KEK) and Japan Aerospace Exploration Agency (JAXA) in Japan, and the K.~A.~Wallenberg Foundation, the Swedish Research Council and the Swedish National Space Board in Sweden. Additional support for science analysis during the operations phase is gratefully acknowledged from the Istituto Nazionale di Astrofisica in Italy and the Centre National d'\'Etudes Spatiales in France. This work performed in part under DOE Contract DE-AC02-76SF00515.

We also want to thank all the observatories from which we used data. We thank the Las Cumbres Observatory and its staff for their continuing support of the ASAS-SN project. ASAS-SN is supported by the Gordon and Betty Moore Foundation through grant GBMF5490 to the Ohio State University, and NSF grants AST-1515927 and AST-1908570. Development of ASAS-SN has been supported by NSF grant AST-0908816, the Mt. Cuba Astronomical Foundation, the Center for Cosmology and AstroParticle Physics at the Ohio State University, the Chinese Academy of Sciences South America Center for Astronomy (CAS-SACA), the Villum Foundation, and George Skestos. The AAVSO database: Kafka, S., 2021, Observations from the AAVSO International Database, \url{https://www.aavso.org}. The CSS survey is funded by the National Aeronautics and Space Administration under Grant No. NNG05GF22G was issued through the Science Mission Directorate Near-Earth Objects Observations Program. The Catalina Real-Time Transient Survey is supported by the U.S.~National Science Foundation under grants AST-0909182 and AST-1313422. This paper has made use of up-to-date SMARTS optical/near-infrared light curves that are available at \url{www.astro.yale.edu/smarts/glast/home.php}. Data from the Steward Observatory spectropolarimetric monitoring project were used. This program is supported by Fermi Guest Investigator grants NNX08AW56G, NNX09AU10G, NNX12AO93G, and NNX15AU81G. This research has made use of data from the OVRO 40-m monitoring program \citep{ovro_monitoring}, supported by private funding from the California Institute of Technology and the Max Planck Institute for Radio Astronomy, and by NASA grants NNX08AW31G, NNX11A043G, and NNX14AQ89G and NSF grants AST-0808050 and AST- 1109911. We also acknowledge the use of public data from the {\it Swift} data archive.

\section{Data Availability}

All the data used in this work are publicly available or available on request to the responsible for the corresponding observatory/facility. All the links to the databases, online repositories, and/or contact information are provided in the footnotes in Section \ref{sec:wave_data}.



\bibliographystyle{mnras}
\bibliography{literature} 




\appendix
\renewcommand{\thefigure}{A\arabic{figure}}
\section*{Appendix}\label{sec:appendix}
This section includes the MWL LCs of the blazars analyzed in this study. 
\begin{figure}
	\centering
	\includegraphics[width=\columnwidth]{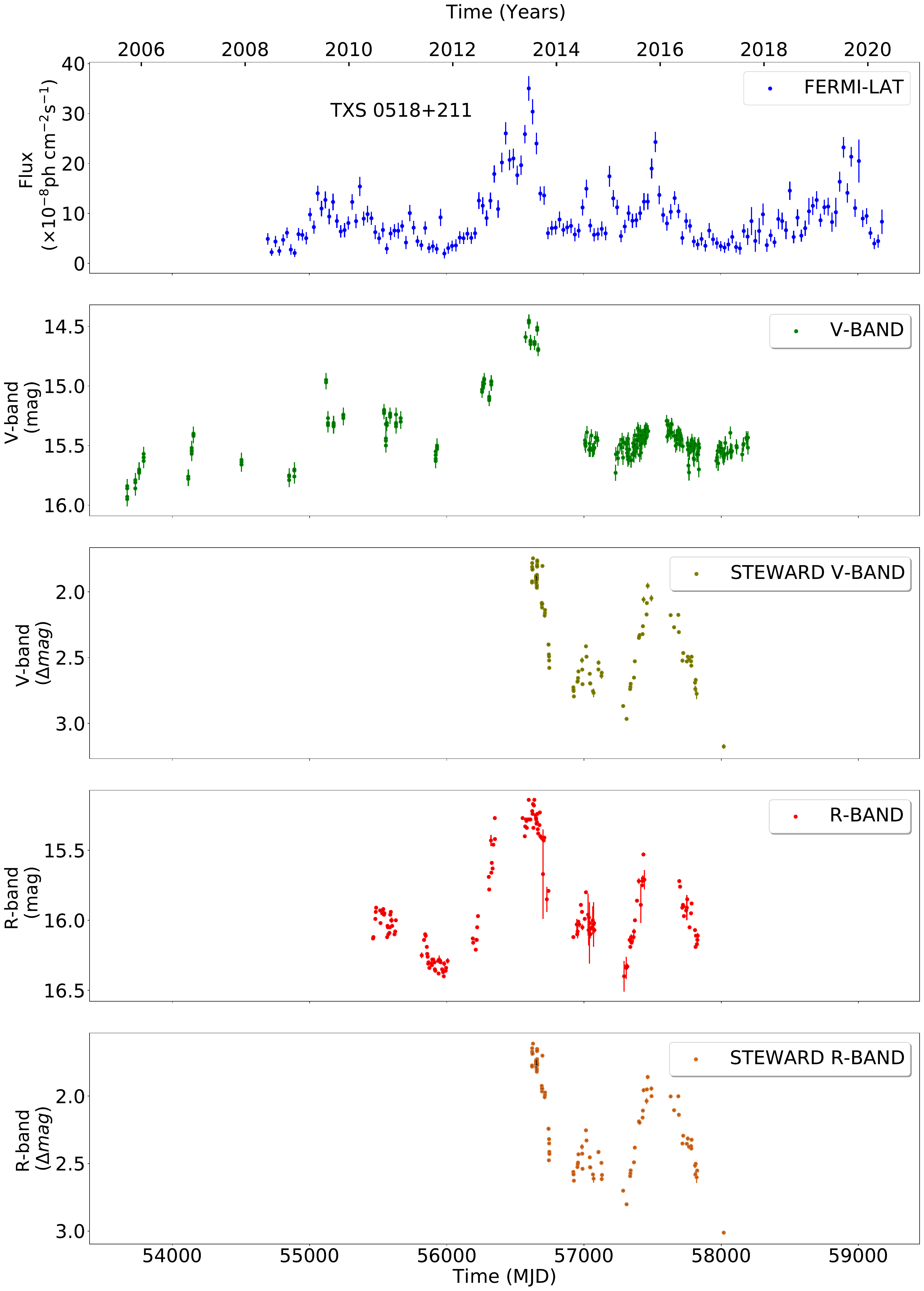}
	\caption{MWL light curves of TXS~0518+211. From top to bottom: \textit{Fermi}-LAT (E $>$ 0.1 GeV), V-band (combination of CSS and ASAS-SN), non-calibrated V-band (Steward Observatory), R-band (KAIT) and non-calibrated R-band (Steward Observatory) light curves.}
\end{figure}

\begin{figure}
	\centering
	\includegraphics[width=\columnwidth]{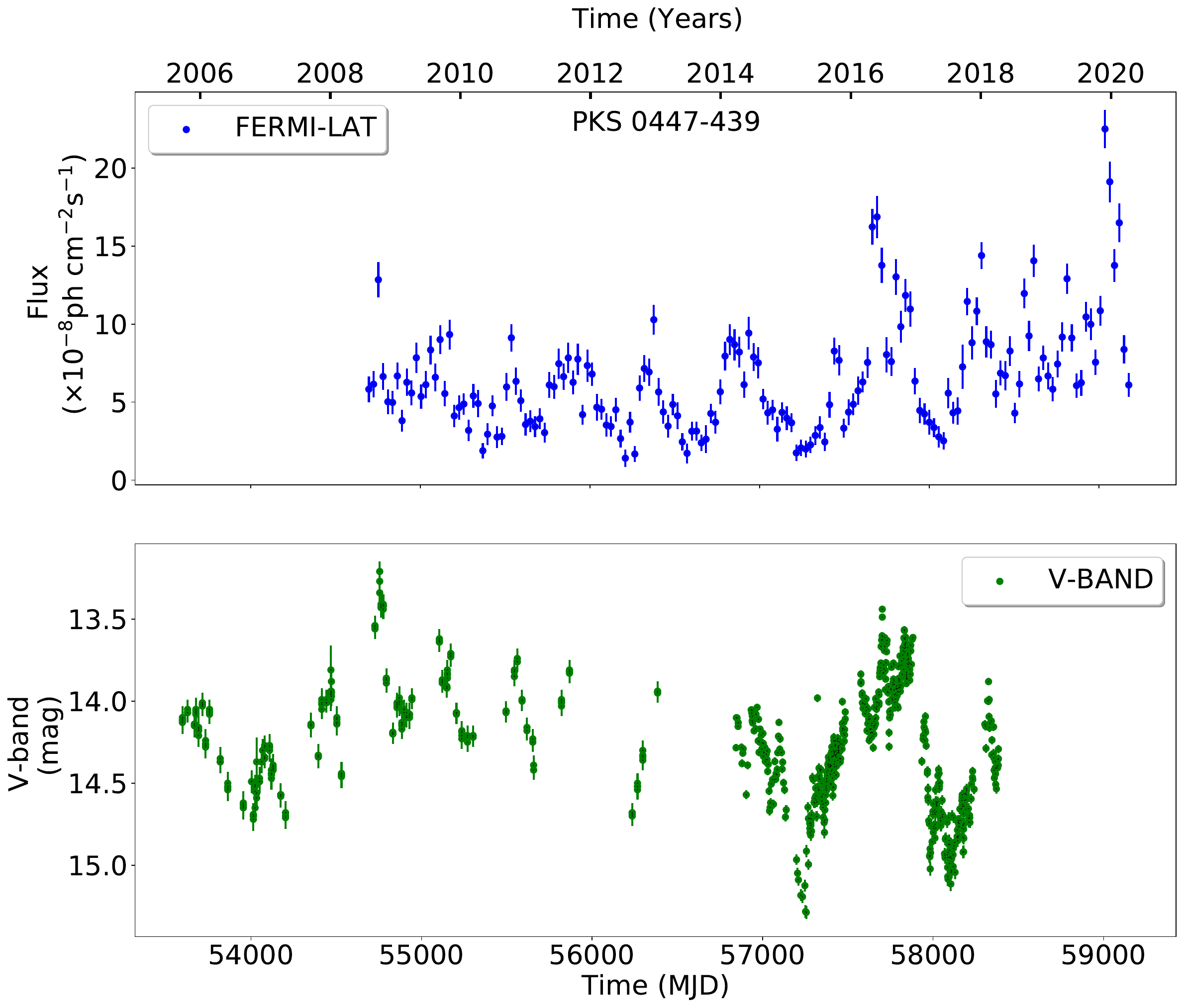}
	\caption{MWL light curves of PKS~0447-439. From top to bottom: \textit{Fermi}-LAT (E $>$ 0.1 GeV) and V-band light curves (combination of CCS and ASAS-SN).}
\end{figure}

\begin{figure}
	\centering
	\includegraphics[width=\columnwidth]{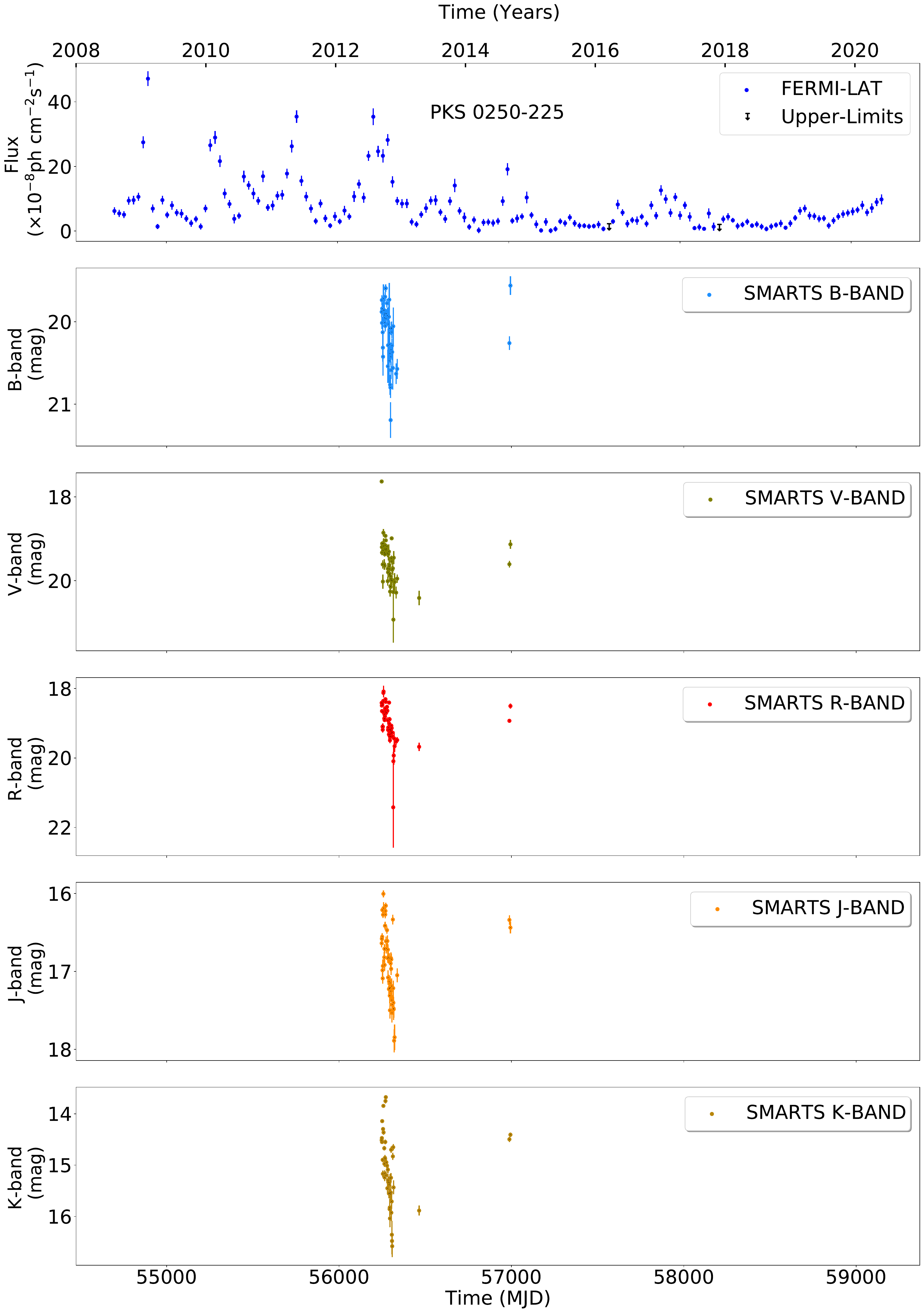}
	\caption{MWL light curves of PKS~0250-225. From top to bottom: \textit{Fermi}-LAT (E $>$ 0.1 GeV), B-band, V-band, R-band, and J-band light curves from SMARTS.}
\end{figure}

\begin{figure}
	\centering
	\includegraphics[width=\columnwidth]{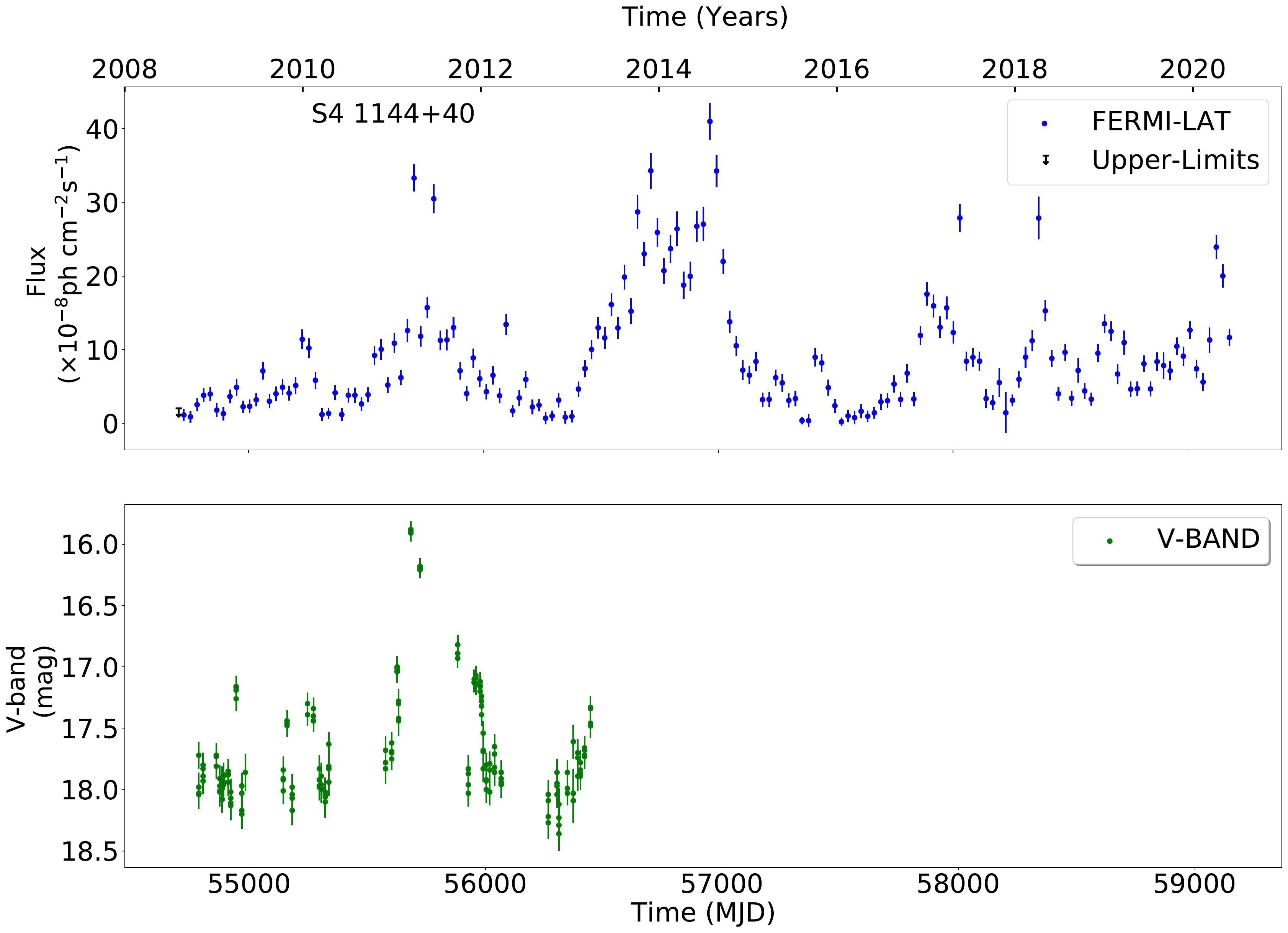}
	\caption{MWL light curves of S4~1144+40. From top to bottom: \textit{Fermi}-LAT (E $>$ 0.1 GeV), and V-band (CSS) light curves.}
\end{figure}

\begin{figure}
	\centering
	\includegraphics[width=\columnwidth]{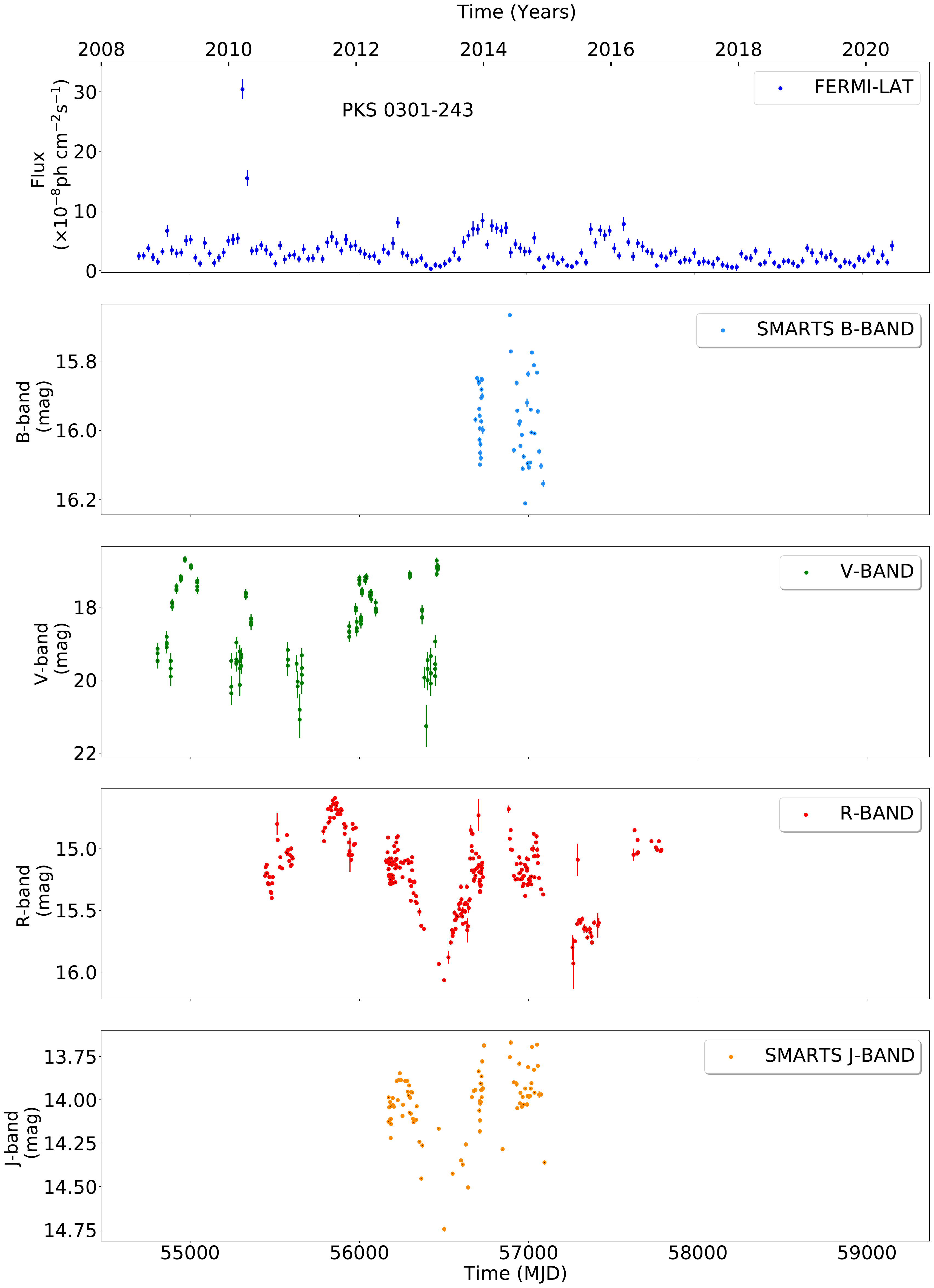}
	\caption{MWL light curves of PKS~0301-243. From top to bottom: \textit{Fermi}-LAT (E $>$ 0.1 GeV), B-band (SMARTS), V-band (CSS and SMARTS), R-band (combination of KAIT and SMARTS), and J-band (SMARTS) light curves.}
\end{figure}

\begin{figure}
	\centering
	\includegraphics[width=\columnwidth]{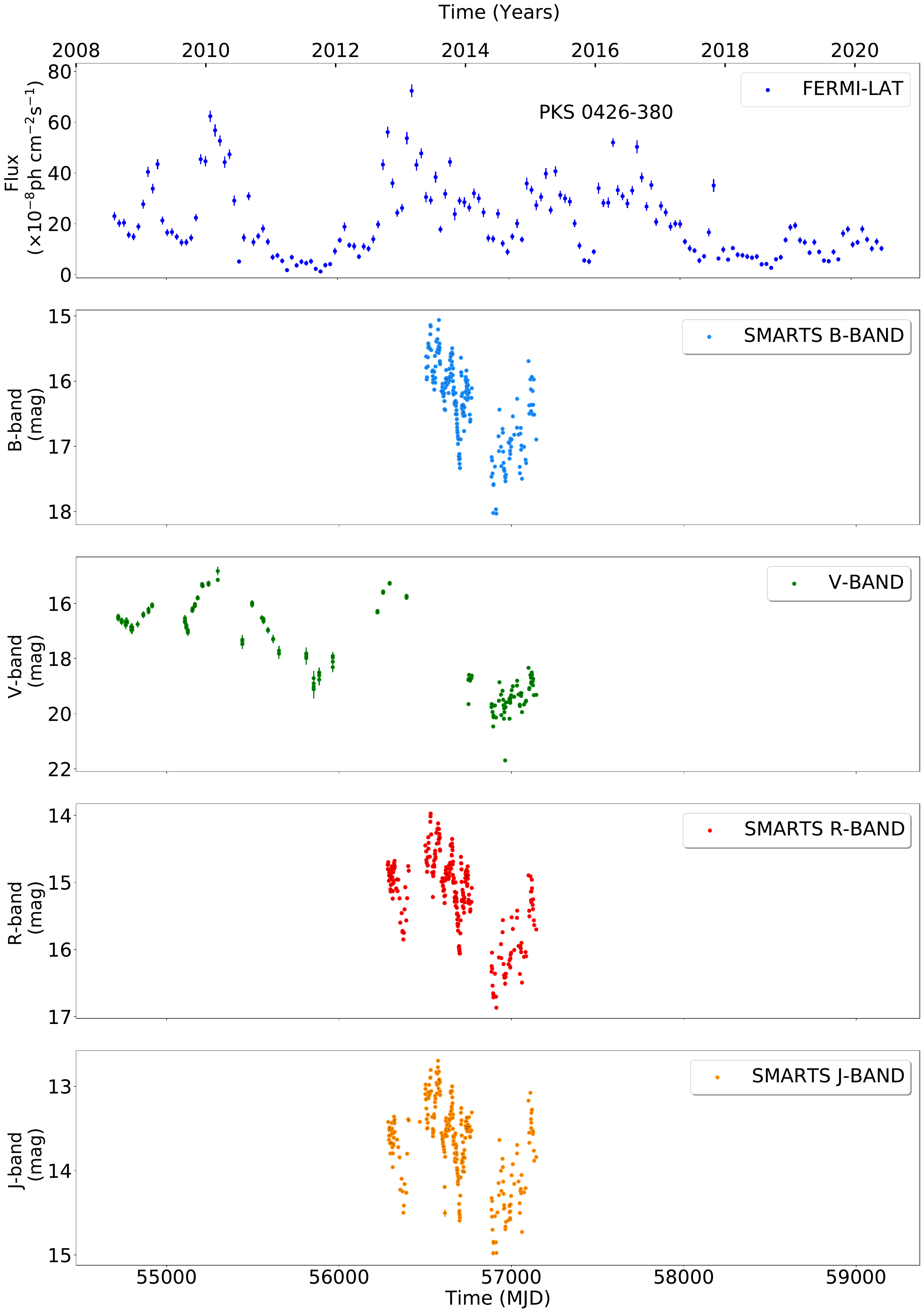}
	\caption{MWL light curves of PKS 0426$-$380. From top to bottom: \textit{Fermi}-LAT (E $>$ 0.1 GeV), B-band (SMARTS), V-band (combination of CSS and SMARTS), R-band (SMARTS), and J-band (SMARTS) light curves.}
\end{figure}

\begin{figure}
	\centering
	\includegraphics[width=\columnwidth]{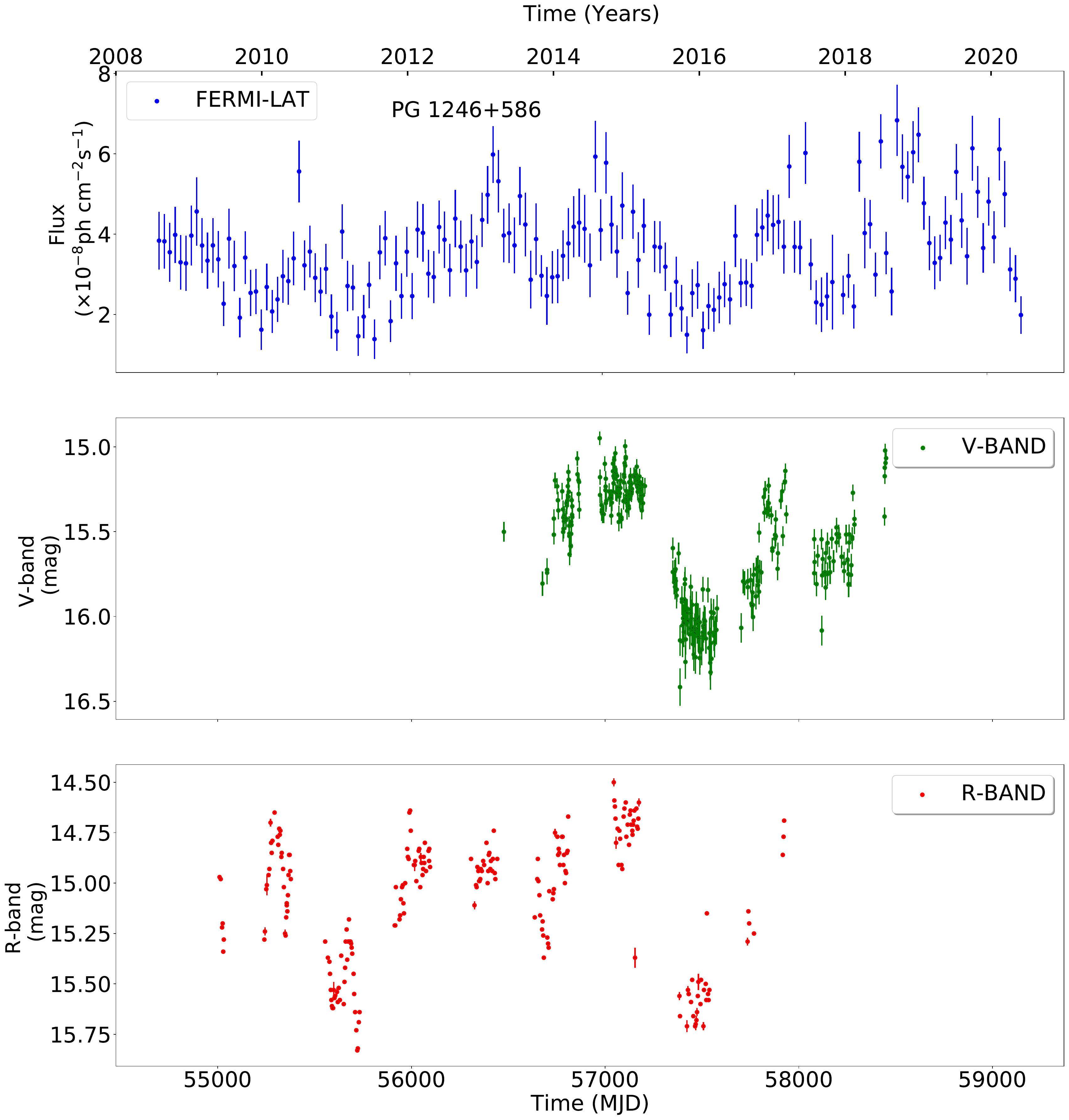}
	\caption{MWL light curves of PG~1246+586. From top to bottom: \textit{Fermi}-LAT (E $>$ 0.1 GeV), V-band (ASAS-SN), and R-band (KAIT) light curves.}
\end{figure}

\begin{figure}
	\centering
	\includegraphics[width=\columnwidth]{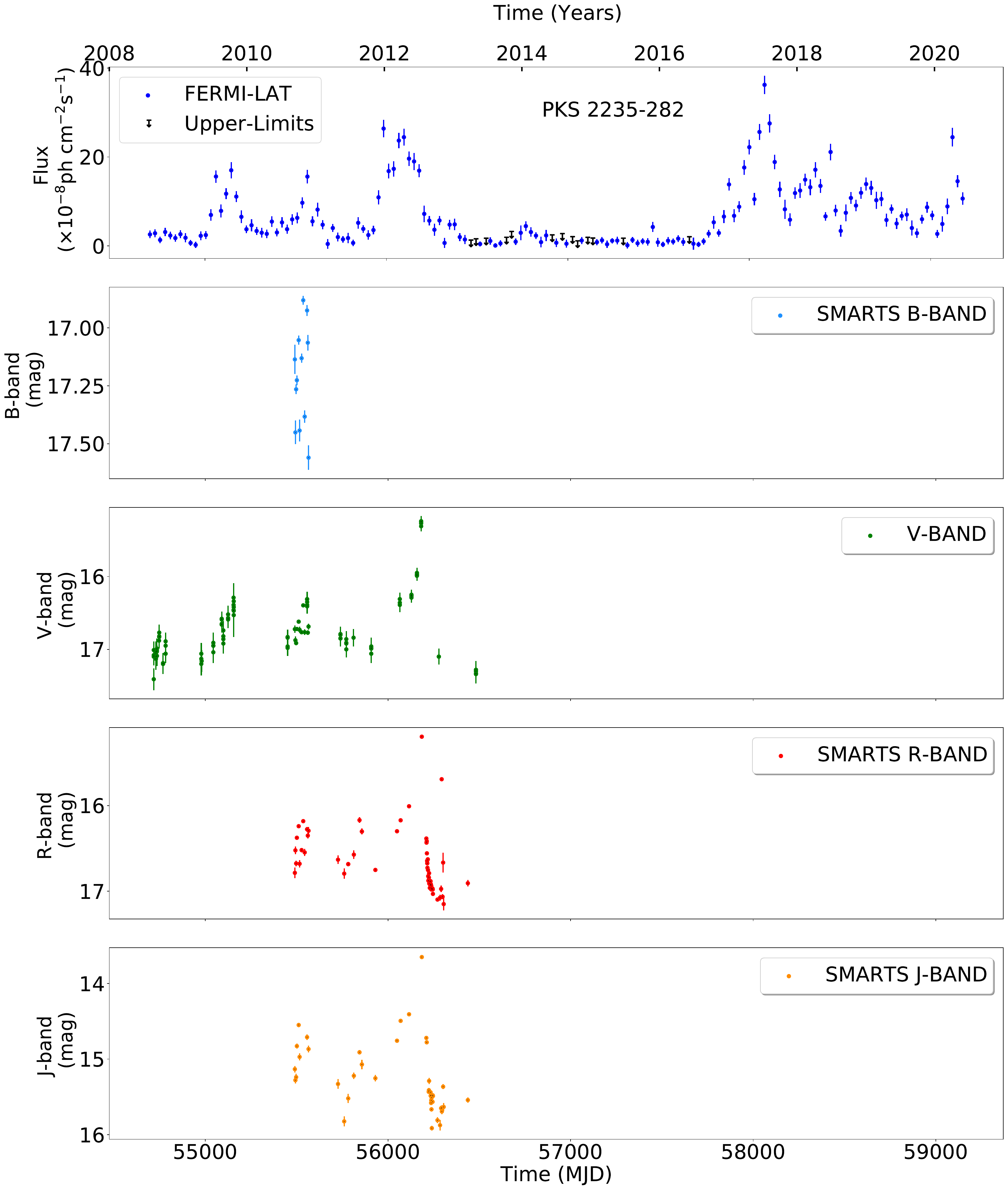}
	\caption{MWL light curves of PKS~2255-282. From top to bottom: \textit{Fermi}-LAT (E $>$ 0.1 GeV), B-band (SMARTS), V-band (combination of CSS and SMARTS), R-band (SMARTS), and J-band (SMARTS) light curves.}
\end{figure}

\begin{figure}
	\centering
	\includegraphics[width=\columnwidth]{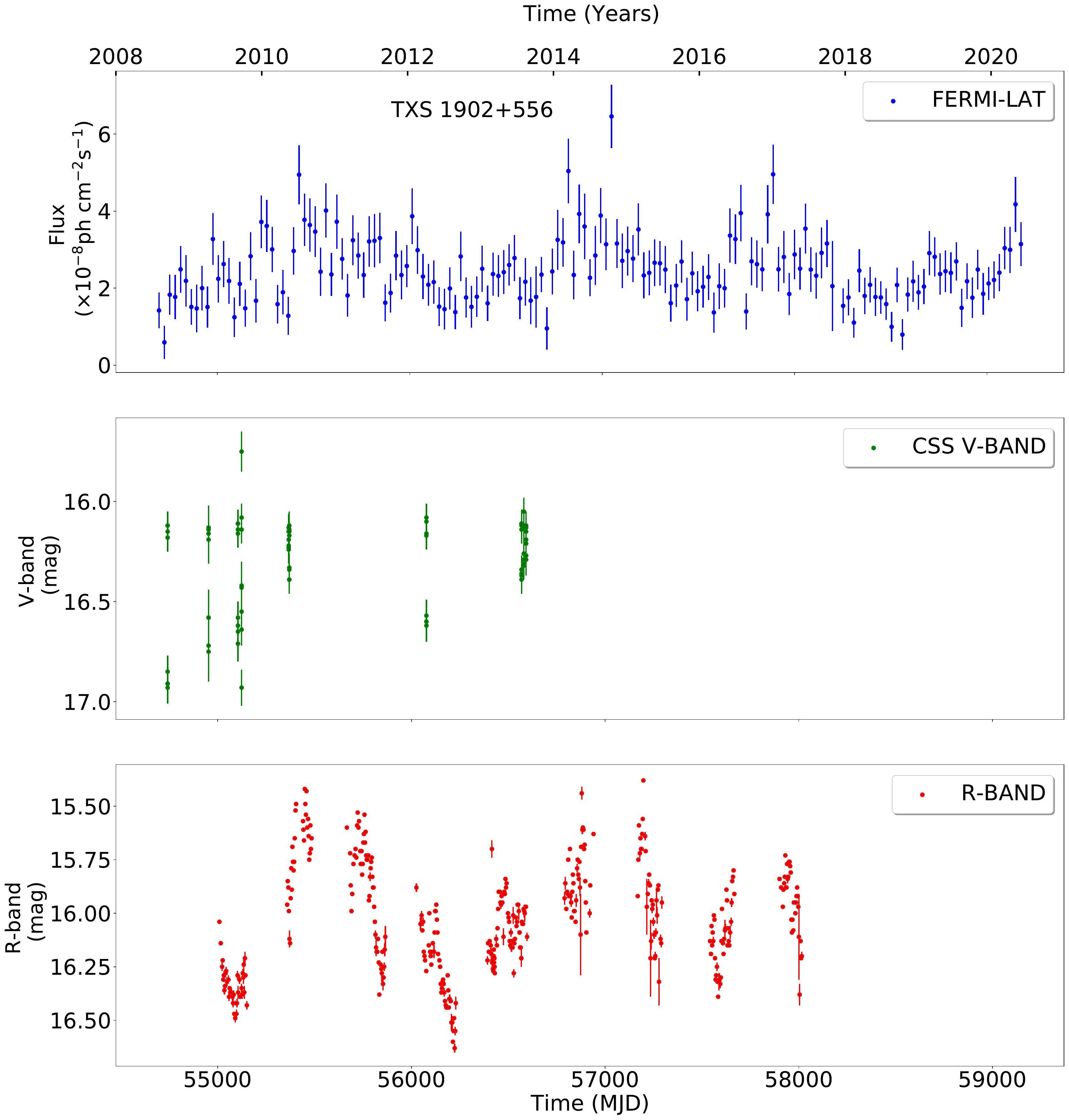}
	\caption{MWL light curves of TXS~1902+556. From top to bottom: \textit{Fermi}-LAT (E $>$ 0.1 GeV), and R-band (KAIT) light curves.}
\end{figure}

\begin{figure}
	\centering
	\includegraphics[width=\columnwidth]{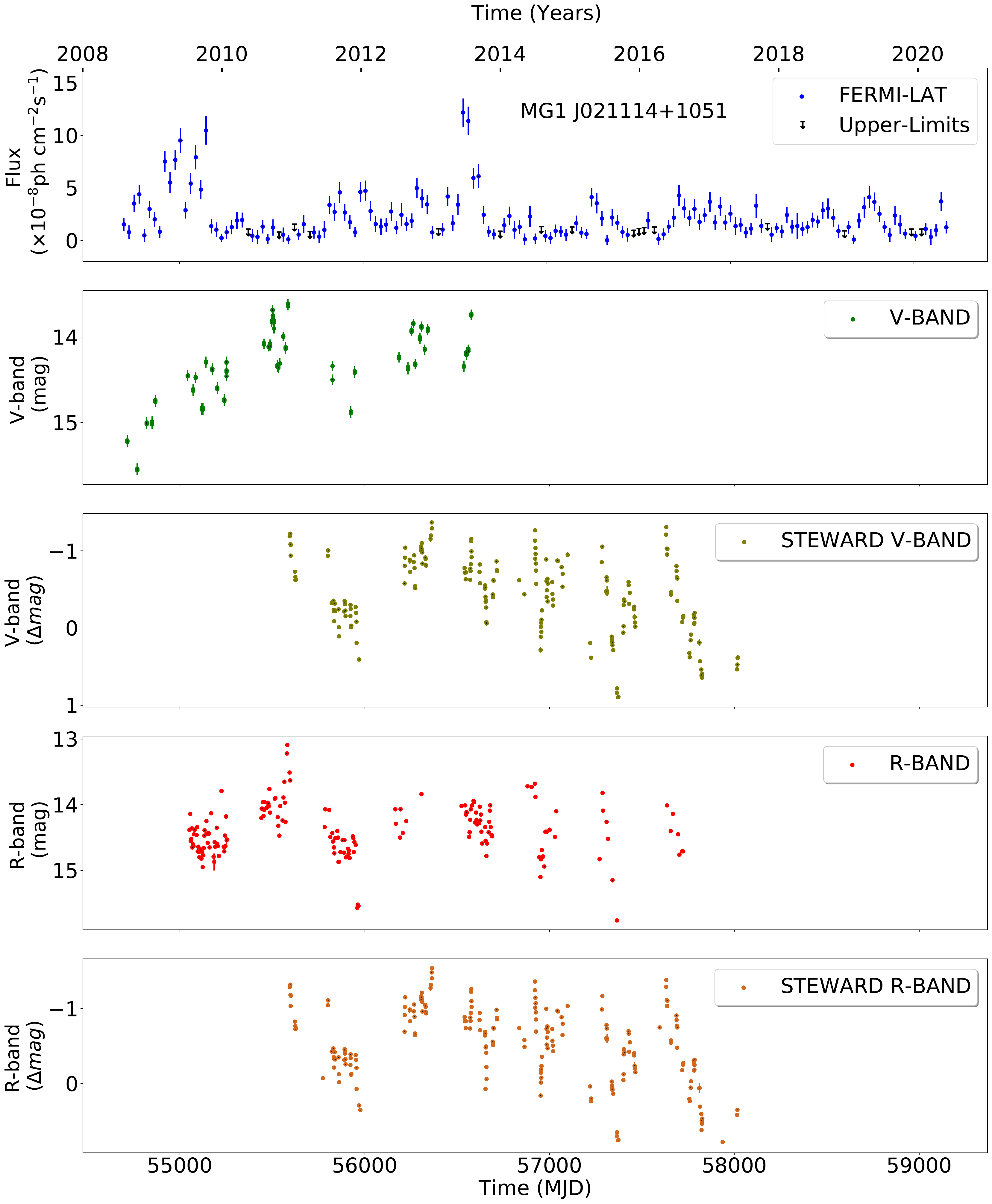}
	\caption{MWL light curves of MG1~J021114+1051. From top to bottom: \textit{Fermi}-LAT (E $>$ 0.1 GeV), V-band (CSS), non-calibrated V-band (Steward Observatory), R-band (KAIT), and non-calibrated R-band (Steward Observatory) light curves.}
\end{figure}

\begin{figure}
	\centering
	\includegraphics[width=\columnwidth]{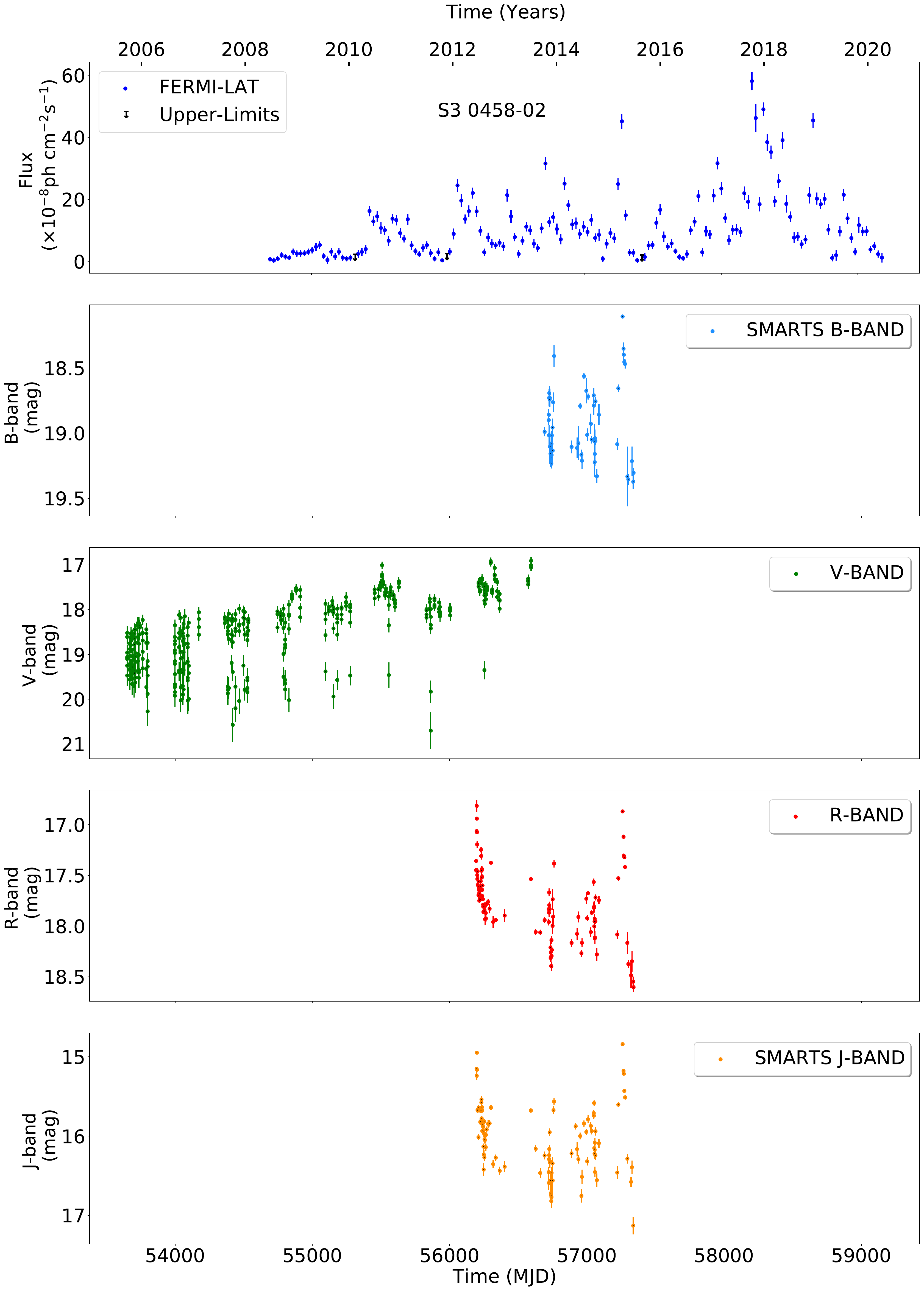}
	\caption{MWL light curves of S3~0458-02. From top to bottom: \textit{Fermi}-LAT (E $>$ 0.1 GeV), B-band (SMARTS), V-band (CSS AMD SMARTS), R-band (SMARTS), and J-band (SMARTS) light curves.}
\end{figure}

\begin{figure}
	\centering
	\includegraphics[width=\columnwidth]{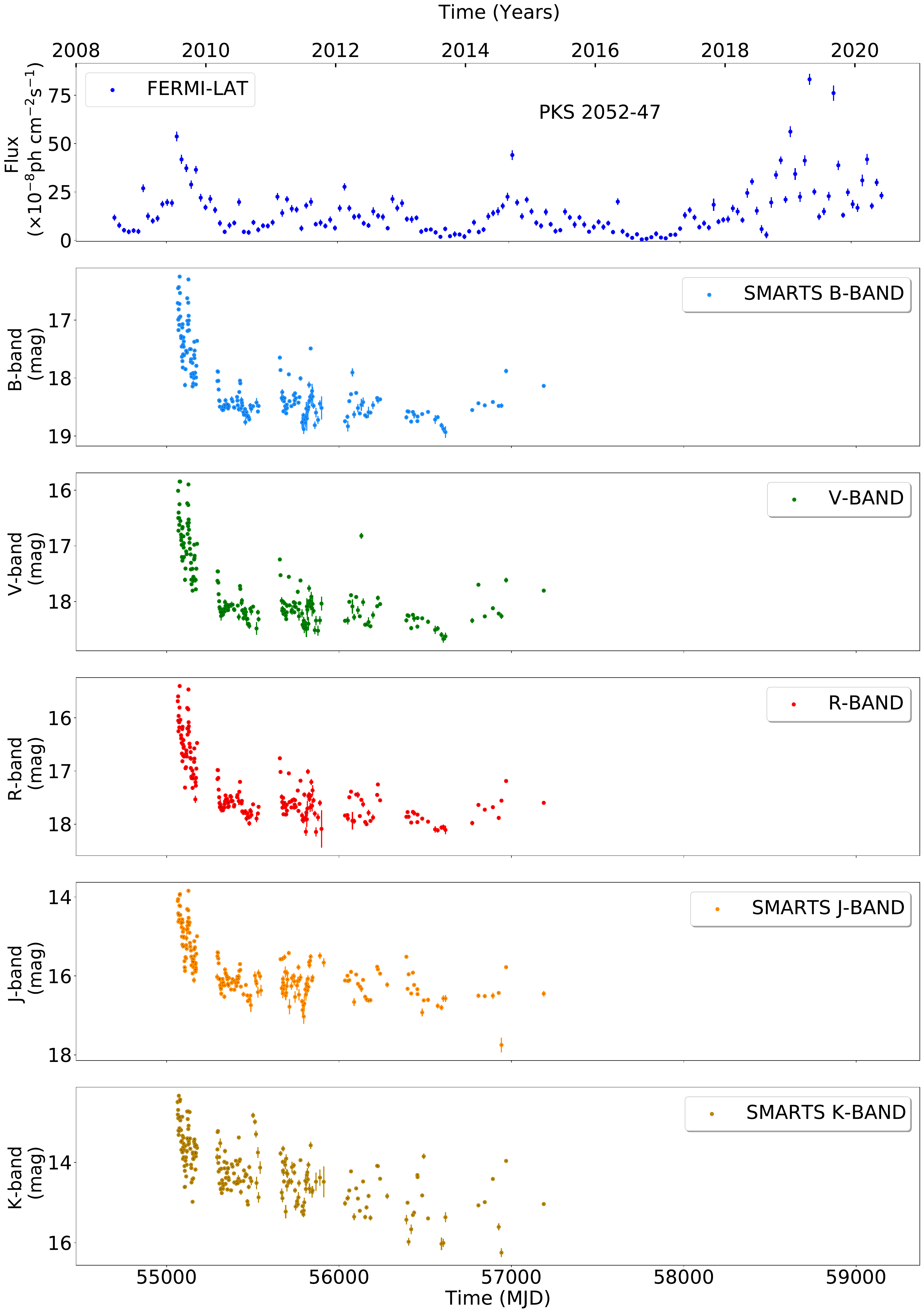}
	\caption{MWL light curves of PKS~2052-47. From top to bottom: \textit{Fermi}-LAT (E $>$ 0.1 GeV),  B-band, V-band, R-band, J-band, and K-band light curves from SMARTS.}
\end{figure}

\begin{figure}
	\centering
	\includegraphics[width=\columnwidth]{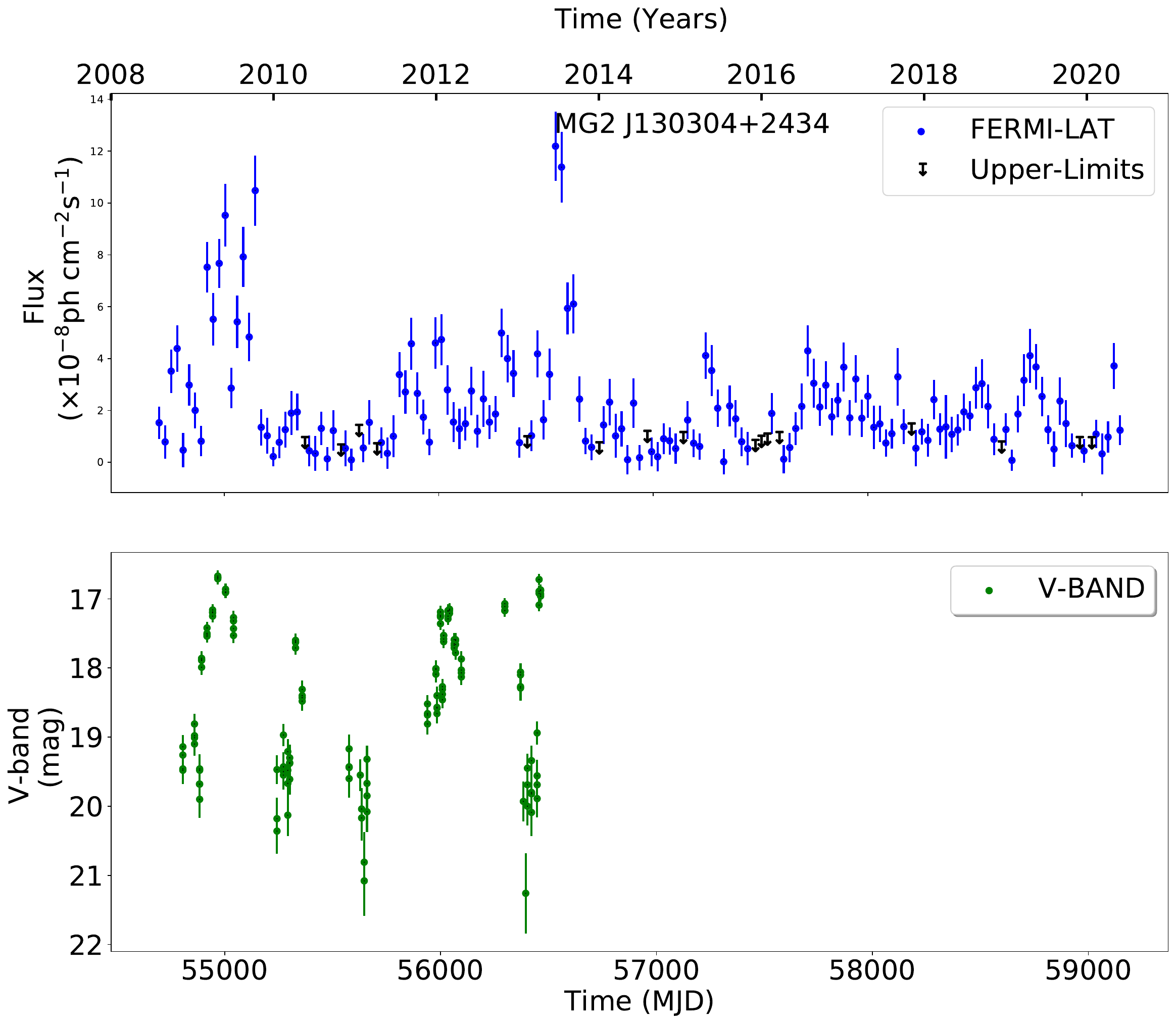}
	\caption{MWL light curves of MG2~J130304+5240236. From top to bottom: \textit{Fermi}-LAT (E $>$ 0.1 GeV) and V-band (CSS) light curves.}
\end{figure}

\begin{figure}
	\centering
	\includegraphics[width=\columnwidth]{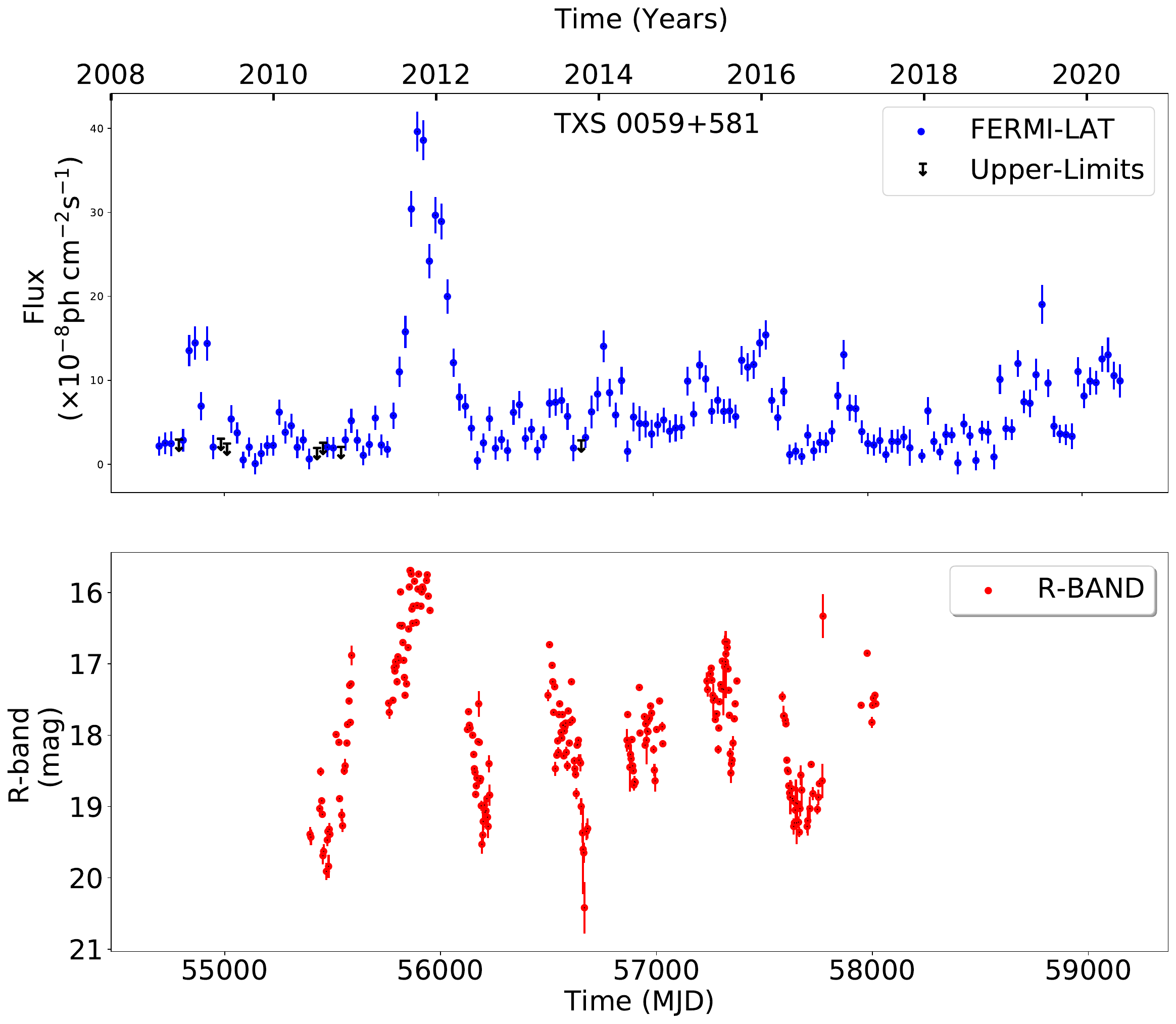}
	\caption{MWL light curves of TXS~0059+581. From top to bottom: \textit{Fermi}-LAT (E $>$ 0.1 GeV) and R-band (KAIT) light curves.}
\end{figure}

\begin{figure}
	\centering
	\includegraphics[width=\columnwidth]{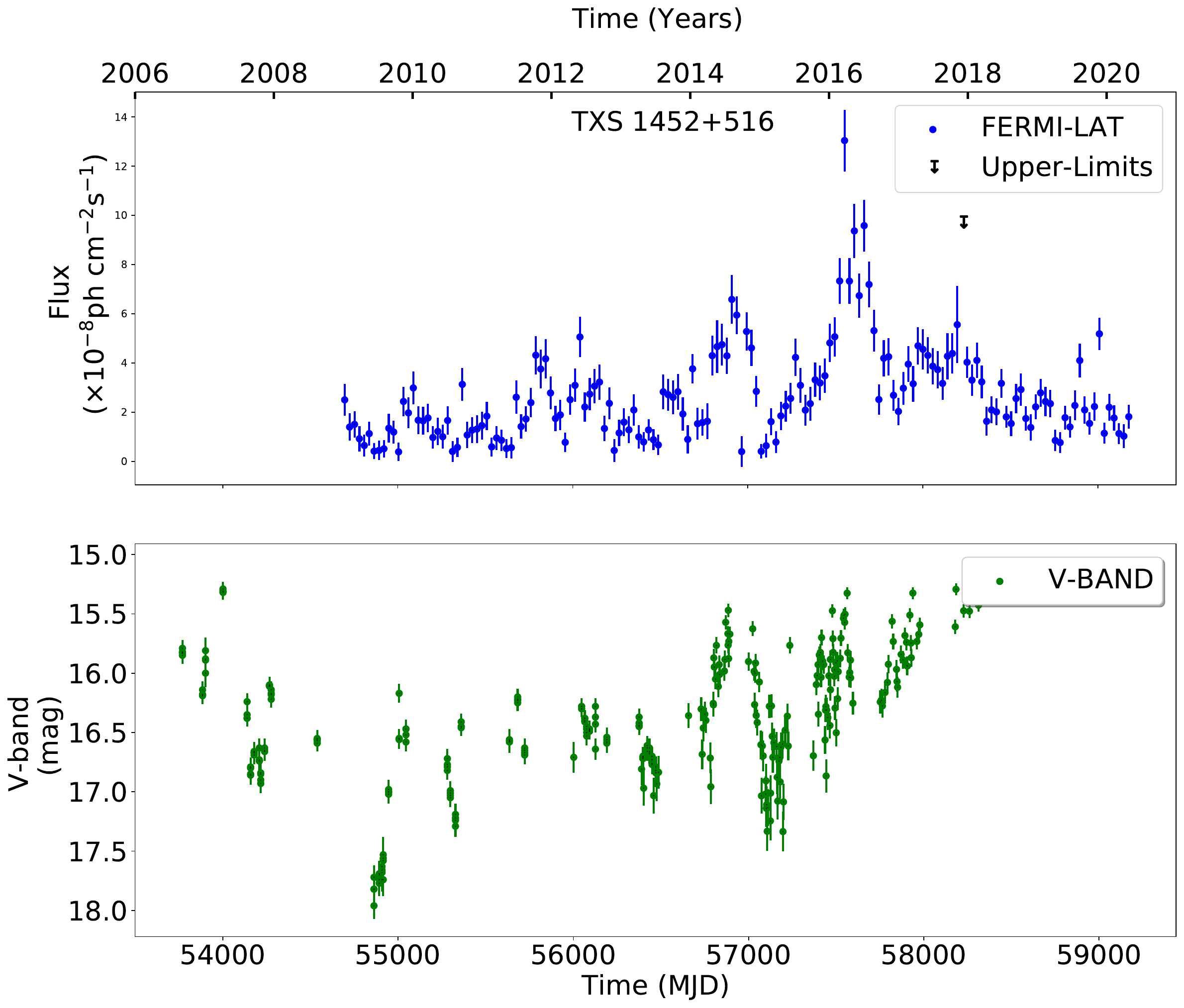}
	\caption{MWL light curves of TXS~1452+516. From top to bottom: \textit{Fermi}-LAT (E $>$ 0.1 GeV) and V-band (combination of CSS and ASAS-SN) light curves.}
\end{figure}

\begin{figure*}
	\centering
	\includegraphics[scale=0.23]{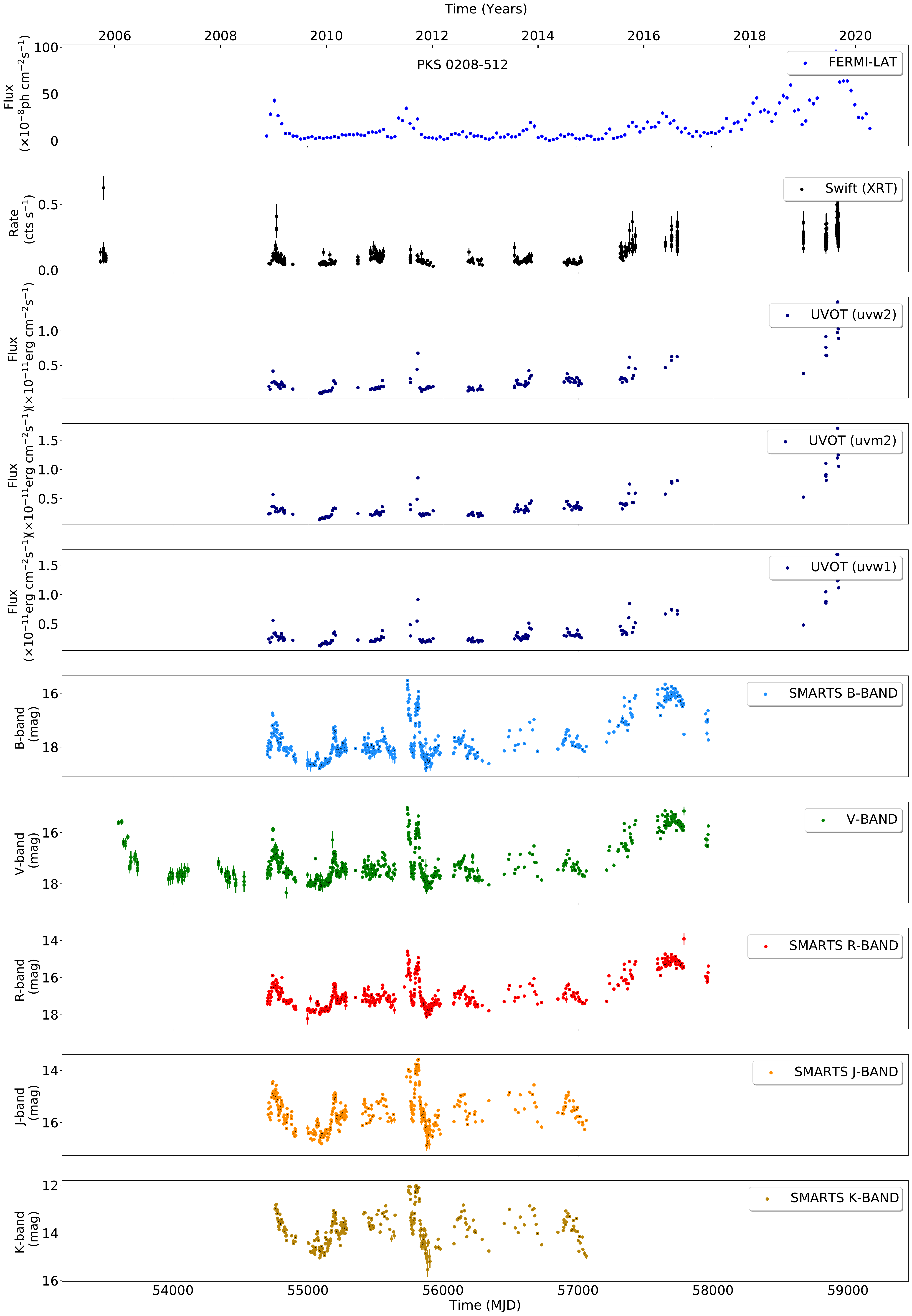}
	\caption{MWL light curves of PKS 0208$-$512. From top to bottom: \textit{Fermi}-LAT (E $>$ 0.1 GeV), {\it Swift}-XRT), UVOT (filters 'uvw2', 'uvm2', and 'uvw1'), B-Band (SMARTS), V-band (combination of CSS and SMARTS), R-band (SMARTS), J-band (SMARTS), and K-band (SMARTS) light curves.}
\end{figure*}

\begin{figure}
	\centering
	\includegraphics[width=\columnwidth]{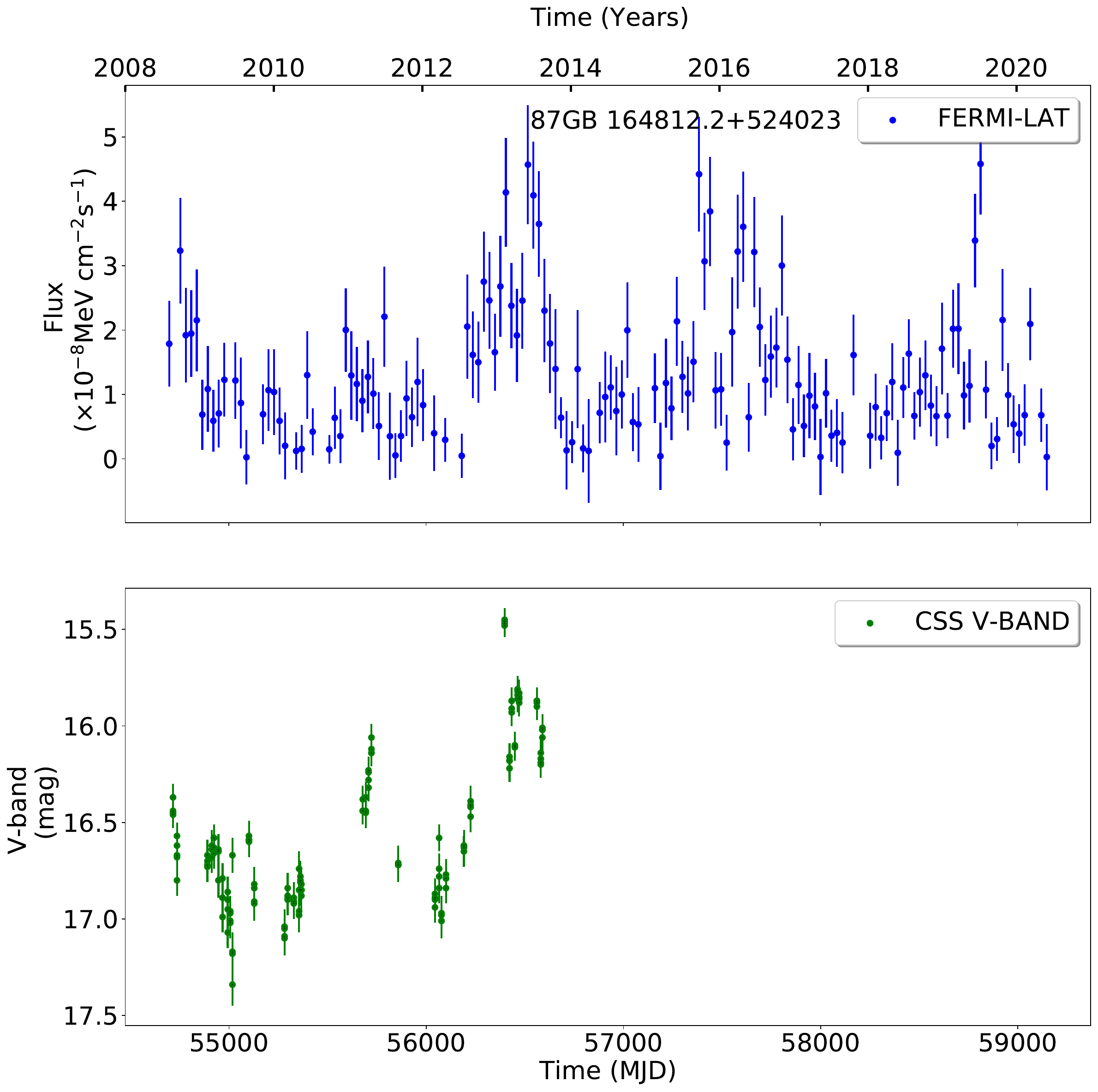}
	\caption{MWL light curves of 87GB~164812.2+524023. From top to bottom: \textit{Fermi}-LAT (E $>$ 0.1 GeV) and V-band (CSS) light curves.}
\end{figure}

\clearpage

\bsp	
\label{lastpage}
\end{document}